\documentclass[fleqn,usenatbib]{mnras}

\usepackage{newtxtext,newtxmath}

\usepackage[T1]{fontenc}

\DeclareRobustCommand{\VAN}[3]{#2}
\let\VANthebibliography\thebibliography
\def\thebibliography{\DeclareRobustCommand{\VAN}[3]{##3}\VANthebibliography}

\usepackage{graphicx}	
\usepackage{amsmath}	

\title[Long baseline ALMA observations of HD\,135344B]{A dusty filament and turbulent CO spirals  in HD\,135344B - SAO\,206462}

\author[S. Casassus et al.]{
Simon Casassus,$^{1}$\thanks{E-mail: simon@das.uchile.cl}
Valentin Christiaens,$^{2}$
Miguel C\'arcamo,$^{3,4,5}$
Sebasti\'an P\'erez,$^{6,5}$
Philipp Weber,$^{1,6}$  \newauthor
Barbara Ercolano,$^{7}$  
Nienke van der Marel,$^{8,9}$ 
Christophe Pinte,$^{2,10}$ 
Ruobing Dong,$^8$
Cl\'ement Baruteau,$^{11}$ \newauthor
Lucas Cieza,$^{12}$
Ewine F. van Dishoeck,$^{13,14}$
Andrés Jordan,$^{15,16}$
Daniel J. Price,$^{2}$ 
Olivier Absil,$^{17,18}$ \newauthor
Carla Arce-Tord,$^{1}$ 
Virginie Faramaz,$^{19}$ 
Christian Flores,$^{20}$ 
Maddalena Reggiani$^{21}$
\\
$^{1}$ Departamento de Astronom\'{\i}a, Universidad de Chile, Casilla 36-D, Santiago, Chile\\
$^{2}$ School of Physics and Astronomy, Monash University, Clayton Vic 3800, Australia\\
$^{3}$ Jodrell Bank Centre for Astrophysics, Department of Physics and Astronomy, University of Manchester,\\ ~~Alan Turing Building, Oxford Road, Manchester, M13 9PL, UK \\
$^{4}$ University of Santiago of Chile (USACH), Faculty of Engineering, Computer Engineering Department, Chile\\
$^{5}$ Center for Interdisciplinary Research in Astrophysics and Space Exploration (CIRAS), Universidad de Santiago de Chile\\
$^{6}$ Departamento de F\'isica, Universidad de Santiago de Chile. Avenida Ecuador 3493, Estaci\'on Central, Santiago, Chile\\
$^{7}$ Universit\"ats-Sternwarte, Ludwig-Maximilians-Universit\"at M\"unchen, Scheinerstr.~1, 81679 M\"unchen, Germany\\
$^{8}$ Department of Physics \& Astronomy, University of Victoria, Victoria, BC, V8P 1A1, Canada\\
$^{9}$ Banting Research fellow\\
$^{10}$ Univ. Grenoble Alpes, CNRS, IPAG, F-38000 Grenoble, France\\
$^{11}$ IRAP, Universit{\'e} de Toulouse, CNRS, UPS, Toulouse, France\\
$^{12}$ N\'ucleo de Astronom\'ia, Facultad de Ingenier\'ia y Ciencias, Universidad Diego Portales, Av Ej\'ercito 441, Santiago, Chile\\
$^{13}$ Leiden Observatory, Leiden University, PO Box 9513, 2300 RA Leiden, The Netherlands\\
$^{14}$ Max-Plank-Institut fur Extraterrestrische Physik, Giessenbachstrasse 1, D-85748 Garching,Germany\\
$^{15}$ Facultad de Ingenier\'ia y Ciencias, Universidad Adolfo Ib\'a\~nez, Av.\ Diagonal las Torres 2640, Pe\~nalol\'en, Santiago, Chile\\
$^{16}$ Millennium Institute for Astrophysics, Chile\\
$^{17}$ STAR Institute, Universit\'e de Li\`ege, 19c All\'ee du Six Ao\^ut, 4000 Li\`ege, Belgium\\
$^{18}$ F.R.S.-FNRS Research Associate\\
$^{19}$ Jet Propulsion Laboratory, California Institute of Technology, 4800 Oak Grove drive, Pasadena CA 91109, USA\\
$^{20}$ Institute for Astronomy, University of Hawaii at Manoa, 640 N. Aohoku Place, Hilo, HI 96720, USA\\
$^{21}$ Institute of Astronomy, KU Leuven, Celestijnenlaan 200D, B-3001 Leuven, Belgium. 
}

\date{Accepted XXX. Received YYY; in original form ZZZ}

\pubyear{2015}

\begin{document}
\label{firstpage}
\pagerange{\pageref{firstpage}--\pageref{lastpage}}
\maketitle

\begin{abstract}
  Planet-disc interactions build up local pressure maxima that may
  halt the radial drift of protoplanetary dust, and pile it up in
  rings and crescents. ALMA observations of the HD\,135344B disc
  revealed two rings in the thermal continuum stemming from
  $\sim$mm-sized dust. At higher frequencies the inner ring is
  brighter relative to the outer ring, which is also shaped as a
  crescent rather than a full ring. In near-IR scattered light images,
  the disc is modulated by a 2-armed grand-design spiral originating
  inside the ALMA inner ring. Such structures may be induced by a
  massive companion evacuating the central cavity, and by a giant
  planet in the gap separating both rings, that channels the accretion
  of small dust and gas through its filamentary wakes while stopping
  the larger dust from crossing the gap.  Here we present ALMA
  observations in the $ J=(2-1)$ CO isotopologue lines and in the
  adjacent continuum, with up to 12\,km baselines. Angular resolutions
  of $\sim 0\farcs03$ reveal the tentative detection of a filament
  connecting both rings, and which coincides with a local
  discontinuity in the pitch angle of the IR spiral, proposed
  previously as the location of the protoplanet driving this
  spiral. Line diagnostics suggest that turbulence, or superposed
  velocity components, is particularly strong in the spirals. The
  $^{12}$CO(2-1) 3-D rotation curve points at stellocentric accretion
  at radii within the inner dust ring, with a radial velocity of up to
  $\sim 5\%\pm 0.5\%$ Keplerian, which corresponds to an excessively
  large accretion rate of $\sim 2\times10^{-6}\,M_\odot\,$yr$^{-1}$ if
  all of the CO layer follows the $^{12}$CO(2-1) kinematics. This
  suggests that only the surface layers of the disc are undergoing
  accretion, and that the line broadening is due to superposed laminar
  flows.
\end{abstract}




\section{Introduction}


The radial drift of protoplanetary dust halts at local pressure maxima
\citep[][]{Weidenschilling1977MNRAS.180...57W}, where the azimuthal
gaseous flow exerts no net drag. The pile-up of dust with
dimensionless stopping time (Stokes number) $S_t \lesssim 1$ in radial
pressure bumps, whichever their origin, explains the Atacama Large
Millimeter/submillimeter Array (ALMA) observations of dusty ringed systems, such as
reported in HLTau \citep[][]{ALMA2015ApJ...808L...3A,
  Carrasco-Gonzalez2019ApJ...883...71C}, in HD\,169142
\citep[e.g.][]{Perez2019AJ....158...15P, Sierra2019ApJ...876....7S},
or in the DSHARP survey \citep[][]{Andrews2018ApJ...869L..41A,
  Dullemond2018ApJ...869L..46D}.





The same radial pressure discontinuities leading to radial trapping
can also trigger the Rossby-wave instability, and develop a large
scale anti-cyclonic vortex
\citep[][]{Lovelace1999ApJ...513..805L,Li2001ApJ...551..874L} resulting in strong
radial and azimuthal concentration for $S_t \lesssim 1$ dust grains
\citep[][]{Birnstiel2013A&A...550L...8B, LyraLin2013ApJ...775...17L,
  Zhu_Stone_2014ApJ...795...53Z, Mittal2015ApJ...798L..25M,
  BZ2016MNRAS.458.3927B}. The radial pressure discontinuity itself
could result from the formation of a planetary gap
\citep[][]{Zhu_Stone_2014ApJ...795...53Z, Koller2003ApJ...596L..91K,
  deVal-Borro2007A&A...471.1043D, ZB2016MNRAS.458.3918Z}, among other
possibilities \citep[e.g.][]{Varniere2006A&A...446L..13V,
  Regaly2012MNRAS.419.1701R}.



Azimuthal dust traps have been identified observationally with ALMA in
the form of large-scale crescents of continuum sub-mm emission, with
extreme azimuthal contrast ratios of $\sim$30 in HD\,142527
\citep[][]{Casassus2013Natur, Casassus2015ApJ...812..126C,
  Muto2015PASJ...67..122M, Boehler2017ApJ...840...60B} and $\sim$\,100
in IRS\,48
\citep[][]{vanderMarel2013Sci...340.1199V,vandermarel_2015ApJ...810L...7V,
  Ohashi2020ApJ...900...81O}.  Such extreme lopsidedness, in
combination with an otherwise full gas disc as revealed by CO
observations, has been interpreted as likely due to dust trapping in a
vortex
\citep[][]{Birnstiel2013A&A...550L...8B,LyraLin2013ApJ...775...17L,
  BZ2016MNRAS.458.3927B, Sierra2017ApJ...850..115S}. Finer ALMA
angular resolutions have revealed that crescents with varying contrast
ratios are fairly frequent in the outer rings of ringed systems, such
as in LkH$\alpha$330 \citep[][]{Isella2013ApJ...775...30I}, SR\,21,
HD135344B \citep[][]{Perez_L_2014ApJ...783L..13P,
  vandermarel_2015A&A...579A.106V, vdM2016ApJ...832..178V,
  Cazzoletti2018A&A...619A.161C}, DoAr\,44
\citep[][]{vdMarel_2016A&A...585A..58V}, and HD\,34282
\citep[][]{vdP2017A&A...607A..55V}. The occurrence of such asymmetries
is associated with Stokes numbers $S_t \lesssim 1$, in agreement with
the dust trapping scenario \citep[e.g.][]{vdM2021AJ....161...33V}. In
the brighter clump of MWC\,758 at VLA frequencies, the multi-frequency
continuum observations can be reproduced with the Lyra-Lin trapping
prescriptions, which yields estimates of the local physical conditions
and constraints on the dust population
\citep[][]{Marino2015ApJ...813...76M, Casassus2019MNRAS.483.3278C}.

Similar processes as in protoplanetary discs are thought to have
occurred in the protosolar disc.  The statistics of meteoritic
inclusions provide important information on the physical processes
that shaped the Solar System.  The same gapped system predicted for an
accreting proto-Jupiter also accounts for the filtering of the larger
dust grains out of the inner disc, and explains the size distributions
of meteoritic inclusions
\citep[][]{Haugbolle2019AJ....158...55H}. Multi-fluid simulations
including dust and gas show that the smaller dust population sieves
across the proto-jovian gap through the planetary wakes, i.e.  as in
the bottle-neck of an hourglass which prevents the larger dust from
crossing the gap, with the maximum grain size permitted to cross
depending on disc characteristics
\citep[e.g.][]{Weber2018ApJ...854..153W}. It is interesting to search
for similar features in exo-protoplanetary systems as required to
explain the meteoritic inclusions: smaller dust grains in the inner
regions of protoplanetary discs hosting deep gaps. Such inner dust
discs are frequently detected, and appear to be depleted in mm-dust
mass \citep[e.g.][]{FrancisvdMarel2020ApJ...892..111F}.


The disc around HD\,135344B, also called SAO\,206462 \citep[with a
  spectral type F8V, and at a distance of
  $135.7\pm1.4$\,pc,][]{Gaia2018A&A...616A...1G}, is particularly
interesting because of several planet-formation signposts, as its
sub-mm continuum emission is essentially composed of two rings, at
$0\farcs4$ (52\,au) and $0\farcs6$ (80\,au) separated by a deep and
broad gap, and crossed by a grand-design 2-armed spiral prominent in
scattered-light images \citep[][]{Stolker2016A&A...595A.113S}. The two
arms of this spiral pattern have been proposed to be driven by two
planets orbiting at $\sim 0\farcs4$ and $\sim 0\farcs9$ \citep[so
  $\sim$54\,au and $\sim$121.5\,au,][]{Muto2012ApJ...748L..22M}, or
from a massive planet inside the cavity at $\sim 0\farcs23$ along with
the outer disc vortex
\citep[][]{vdMarel_2016A&A...585A..58V,vdM2016ApJ...832..178V}, or the
whole pattern could result from gravitational instability
\citep[GI,][]{Dong2015ApJ...812L..32D,
  Dong2018ApJ...862..103D}. Multi-epoch monitoring of the
scattered-light spirals has allowed the measurement of the pattern
motion, which points at a companion at $86^{+18}_{-13}$\,au if the
arms are co-moving, or else to two companions at $49_{-5}^{+6}$\,au and
$120\pm30$\,au \citep[][]{Xie2021ApJ...906L...9X}.  The
multi-wavelength imaging reported by
\citet[][]{Cazzoletti2018A&A...619A.161C} supports an interpretation
in terms of the filtering of the larger dust grains out of the inner
regions, since the inner ring is increasingly brighter at higher
frequencies compared to the outer ring. This brightness effect could
also point to a difference in dust evolution within the two rings, i.e. to 
different fragmentation rates.

Here we report on  new ALMA observations of
HD\,135344B (Section\,\ref{sec:obs}) that reveal the tentative
detection of the dusty protoplanetary wakes predicted by theory. A
narrow trickle of dust continuum emission seems to connect the
lopsided outer ring with the inner ring (Sec.\,\ref{sec:analysis}),
close to the planet at $\sim 0\farcs4$ proposed by
\citet{Muto2012ApJ...748L..22M}. A simple analysis of the line
emission reveals that the 2-armed spiral corresponds to enhanced
velocity dispersions (Sec.\,\ref{sec:analysis}). We interpret the
available data in terms of  disc-surface accretion towards the star 
(Sec.\,\ref{sec:discussion}).

\section{Observations} \label{sec:obs}




HD\,135344B was observed with ALMA during Cycle\,6, as part of program
{\tt 2018.1.01066.S}. The data presented in this article correspond to
a partial delivery of the whole program, which is scheduled for
execution in Cycle\,7 with a nominal 13\,h of telescope time. In this
partial dataset, HD\,135344B was observed from 13-Jul-2019/02:43:39 to
13-Jul-2019/03:55:34 (UTC), for a total of 43\,min on-source. The
phase centre of the array pointed at J2000 15h15m48.4142s
$-37$d09m16.4776s, which is offset by 5.8\,mas from the ICRS position
of the star at the epoch of observations, J2000 15h15m48.4147s
$-$37d09m16.4785s. The array counted 43 active antennas, with
baselines ranging from 111.2\,m to 12.6\,km. The column of
precipitable water vapour (PWV) ranged from 0.9\,mm at the beginning
of the integration, to 0.7\,mm at the end. The correlator was setup to
provide 4 spectral windows (spws): spw\,0, sampling the continuum
around 218\,GHz with 128 channels over a total bandwidth of 2\,GHz;
spw\,1, for $^{12}$CO(2-1) at a rest frequency of 230.538\,GHz,
sampled with 158.74\,m\,s$^{-1}$ channels; spw\,2, sampling the
continuum around 232\,GHz with 960 channels over a total bandwidth of
1.875\,GHz; spw\,3, for C$^{18}$O(2-1) at a rest frequency of
219.560\,GHz, sampled with 166.68\,m\,s$^{-1}$ channels; spw\,4, for
$^{13}$CO(2-1) at a rest frequency of 220.399\,GHz, sampled with
166.04\,m\,s$^{-1}$ channels.  The data were calibrated by staff from
the North America ALMA Regional Center.

Spectral windows spw\,0 and spw\,2 were devoid of conspicuous line
emission and were combined to image the continuum. A single
self-calibration loop, setting up CASA task {\tt gaincal} to average
whole scans (option `solint' set to `inf'), provided an improvement in
{\tt tclean} images using Briggs weights with robustness parameter
$r=0.0$, with a dynamic range (i.e. signal-to-noise ratio) increasing
from $\sim$14.8 to $\sim$17.3. The quantitative improvement was small
but the self-calibration loop got rid of noisy patches, so we adopted
the self-calibrated dataset. We then applied the {\sc uvmem} package
\citep[][]{Casassus2006, Carcamo2018A&C....22...16C} on the
self-calibrated continuum data to produce a non-parametric model
image. In brief, {\sc uvmem} produces a model-image $I^m_j$ and model visibilities $V^m_k$ that fit the
visibility data $V^\circ_k$ in a least-square sense, with the possible inclusion
of a regularising term, by minimising a  figure of merit $L$:
\begin{equation}
  L = \sum_{k=1}^{N_{\rm vis}} \omega_k  |V^\circ_k - V^m_k|^2 + \lambda S \label{eq:L},
\end{equation}
where $\omega_k$ correspond to the visibility weights and the sum runs
over all visibility data (i.e. without gridding). The term $\lambda S$
in Eq.\,\ref{eq:L} is a regularisation term, whose functional
definition depends on the application. In this case image positivity
provided sufficient regularisation, so we restricted the optimisation
to the least-squares term only \citep[see for example ][for example
  regularisation terms and detailed applications of {\sc uvmem} to
  ALMA data in protoplanetary discs]{Casassus2018MNRAS.477.5104C,
  Casassus2019MNRAS.483.3278C}.

\begin{figure*}
  \centering
  \includegraphics[width=\textwidth,height=!]{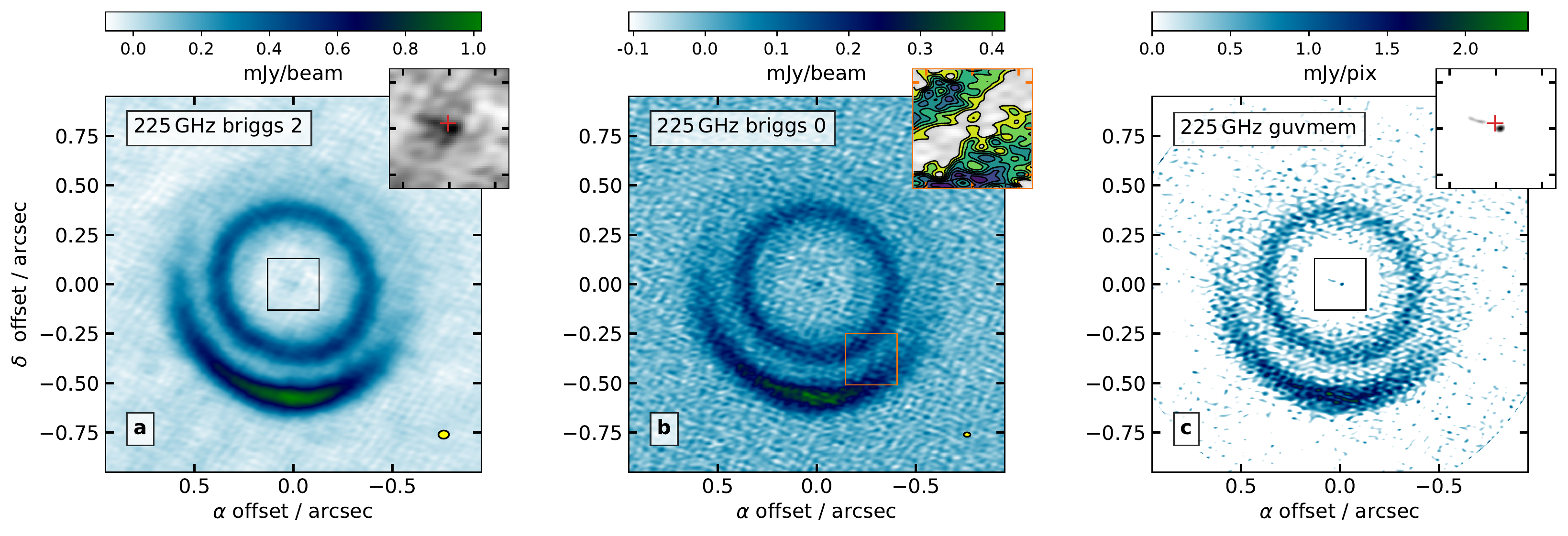}
  \caption{Continuum imaging in HD\,135344B.  {\bf a:} CASA-{\tt
      tclean} image with Briggs robustness parameter $r= 2.0$ (so
    close to natural weights), resulting in a clean beam of
    $0\farcs054\times 0\farcs041 \,/\,-87$\,deg, where we give the
    beam major axis (bmaj), minor axis (bmin) and direction (bpa) in
    the format bmaj$\times$bmin/bpa.  {\bf b:} CASA-{\tt tclean} image
    with Briggs robust parameter 0.0, with a clean beam
    $0\farcs034\times0\farcs022\,/\,88$\,deg {\bf c:} {\sc uvmem}
    model image, with an effective angular resolution of $\sim$1/3 of
    the natural-weights beam, or $\sim 0\farcs018 \times
    0\farcs014\,/\,-87$\,deg.  The insets in a) and c) zoom inside the
    central emission, where the centroid of the inner ring is marked by a
    red symbol (see Sec.\,\ref{sec:analysis_continuum} for
    details). The inset in b) zooms on the filament     with contour
    levels starting at $4\sigma$ and incremented in units of $\sigma$ 
    (the centre of this  inset is at 
    $[-0\farcs276,-0\farcs378]$). The tick-marks in the insets are
    separated by $0\farcs1$.   } \label{fig:HD135344B_cont}
\end{figure*}

As summarised in Fig.\,\ref{fig:HD135344B_cont}, the continuum data are
consistent with the general structure of the disc previously reported
by \citet{Cazzoletti2018A&A...619A.161C}. The disc is composed of two
main features: an outer ring  shaped into a large crescent, as
expected for a vortex \citep[e.g.][]{BZ2016MNRAS.458.3927B}, and an
 inner ring that at this frequency appears fainter relative to the outer ring. Here we also notice a faint plateau or pedestal abut onto
the inner ring, and enclosing an otherwise very deep cavity. The
central source, probably related to the star, shows signal at
5$\sigma$, with intriguing structure the details  of which should be
ascertained in second epoch imaging.

When integrated over the whole field plotted in
Fig.\,\ref{fig:HD135344B_cont}, the continuum flux density at 225\,GHz
from HD\,135344B is $117.4\pm0.1$\,mJy in the Briggs 2.0 image, with a
noise of $\sim 18\,\mu$Jy\,beam$^{-1}$, and $105.1\pm0.3$\,mJy in the
Briggs 0.0 image, with a noise of $\sim 24\,\mu$Jy\,beam$^{-1}$. The
noise in the {\tt uvmem} model image is positive definite, but we
quote the integrated flux density for completeness, of $114.6\pm
1$\,mJy with a noise of $\sim 1\,\mu$\,Jy\,pix$^{-1}$ (corresponding
to the peak signal of the spurious features away from the source). The
thermal uncertainty is very small, but both values are affected by a
systematic uncertainty of $\sim$10\% root-mean-square (rms). The
nominal maximum recoverable angular scale (MRS) for this array
configuration is $\sim$0\farcs62 (as given in the ALMA proposer's
guide for Cycle\,6), and since the continuum signal extends over
$\sim$1\farcs2, there may be a measure of flux loss, i.e. missing low
spatial frequencies in the reconstructed image due to missing
short-spacings in the $uv-$coverage. The slightly lower flux in Briggs
0.0 compared to Briggs 2.0 may perhaps reflect this flux loss
effect. But the total missing flux must be fairly small, as there are
no obvious image synthesis artefacts that modulate the signal.

The line datacubes were extracted following a standard procedure, and
after propagating the self-calibration gain tables to the full
dataset. Continuum subtraction was performed using CASA task {\tt
  uvcontsub}, and using a linear model that avoids the vicinity of the
lines. The data were then resampled into the local standard of rest,
and using spectral channels with a width of 0.2\,km\,s$^{-1}$, common
to all lines. In an initial attempt at imaging synthesis, the resulting
datacubes were first  imaged with CASA task {\tt tclean}, in its
multi-scale version, with automatic masking (as implemented in the
`auto-multithresh' option to {\tt tclean}), and using Briggs $r=2.0$.
In the case of the $^{12}$CO(2-1) line, the peak signal is
$\sim$18\,mJy\,beam$^{-1}$, while the noise (including image synthesis
residuals) is 2.2\,mJy\,beam$^{-1}$ (for a beam $0\farcs054 \times
0\farcs040 / 89.4 $\,deg\footnote{expressed in the form
(BMAJ$\times$BMIN / BPA), where BMAJ and BMIN are the full-width major
and minor axis, and BPA is the beam PA in degrees East of
North.}).

The resulting channel maps are shown in
Appendix\,\ref{sec:channelmaps}. Imaging the $^{12}$CO(2-1) datacubes
is particularly challenging because this preliminary data release did
not include data acquired in a complementary compact
configuration. This results in strong aliasing in the restored images,
as can be assessed by inspection of the channel maps in
Fig.\,\ref{fig:HD135344B_12CO21_channels}.

In an effort to overcome
the systematics due to image synthesis, we also explored a different
strategy using our {\sc uvmem} package. We proceeded as for the continuum emission, but this time including  an
entropic term for regularisation as the data are quite noisy. We minimised $L$ in Eq.\,\ref{eq:L} with 
\begin{equation}
S = \sum_{j=1}^{N_{\rm pix}} \frac{I_j^m}{M} \ln\left(\frac{I_j^m}{M}\right),
\end{equation}
where the sum runs over all pixels in the image (or within a
user-defined region enclosing the signal), and $M$ is the default
pixel intensity value, and is set to $10^{-3}$ times the theoretical
noise of the dirty map (as inferred from the visibility weights). Here
we use a control factor $\lambda = 1\times10^{-3}$. An important
difference between {\sc uvmem} and {\tt tclean} is that {\tt tclean}
uses masks to select regions in the sky which correspond to
signal. These masks are adjusted iteratively in {\tt tclean}.  Initial
trials with blind reconstructions using {\sc uvmem}, i.e. without
applying masks, yielded similar results as for the {\tt tclean}
reconstructions. The restored data cubes, obtained using the {\sc
  uvmem} model datacube and Briggs $r=2$, yielded a similar dynamic
range as with {\tt tclean}, but are less clumpy and more sensitive to
extended signal. However, we also attempted the implementation in {\sc
  uvmem} of masks in channel maps, similar to those used by {\tt
  tclean}. We used a rough approximation to Keplerian masks, using the
tools\footnote{\url{https://github.com/richteague/keplerian_mask}}
developed by \citet{rich_teague_2020_4321137}. The use of masks, as in
{\tt tclean}, resulted in an important improvement in dynamic range,
as can be judged by comparing the {\tt tclean} channel maps with the
{\sc uvmem} channel maps shown in
Fig.\,\ref{fig:HD135344B_12CO21_channels_uvmem}. The data from the
rarer isotopologues are much noisier and compact and their dynamic
range is limited by thermal noise rather than synthesis imaging
artefacts, and indeed {\sc uvmem} did not provide improvements in
their cases.

\begin{figure*}
  \centering
  \includegraphics[width=\textwidth,height=!]{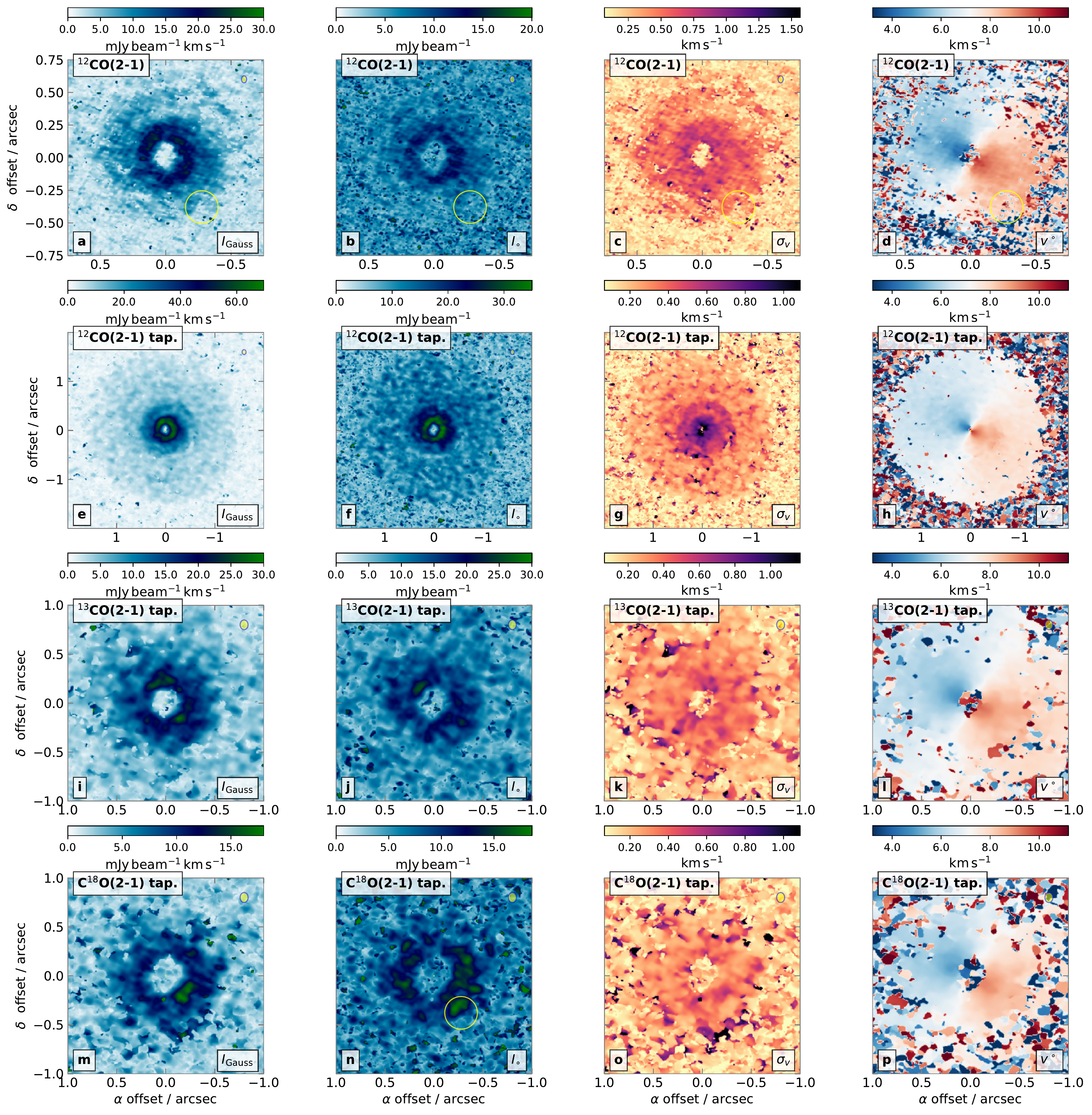}
  \caption{Moment maps in $^{12}$CO(2-1), $^{13}$CO(2-1) and
    C$^{18}$O(2-1) extracted from the {\tt uvmem} datacubes using
    single-Gaussian fits in velocity. The yellow circle marks the same
    region of interest as in Fig.\,\ref{fig:HD135344B_cont}b.  The
    first row shows, for $^{12}$CO(2-1), {\bf a)} the Gaussian
    velocity-integrated intensity $I_{\rm Gauss}$, {\bf b)} the
    Gaussian amplitude $I_\circ$, {\bf c)} the Gaussian dispersion
    $\sigma_v$, and {\bf d)} the Gaussian velocity centroid
    $v_\circ$. The beam ($0\farcs054 \times 0\farcs040 \,/\,0$\,deg)
    is indicated by a yellow ellipse.  The second to fourth rows
    extend the same images to tapered versions of $^{12}$CO(2-1)
    ($0\farcs100\times 0\farcs081 \,/\,0$\,deg beam), $^{13}$CO(2-1)
    ($0\farcs100\times 0\farcs081 \,/\,0$\,deg beam), and
    C$^{18}$O(2-1) ($0\farcs102\times 0\farcs082 \,/\,0$\,deg
    beam). Note the larger field of view for e)-h).
  } \label{fig:sgaussmoms}
\end{figure*}

In order to trace the whole disc and  extend the field of the isotopologue
analysis presented below, we smoothed the data with a circular
Gaussian taper, applied in the $uv$-plane, with a width of
50\,mas. The resulting tapered beam is $\sim 0\farcs1\times0\farcs08 /
0$\,deg (see caption to Figs.\,\ref{fig:sgaussmoms}). These tapered
versions of the data were again imaged using {\sc uvmem} with
Keplerian masks, but without entropic regularisation (so with
$\lambda=0$).



The moments maps shown in Figs.\,\ref{fig:sgaussmoms} were extracted
using single-Gaussian fits to each datacube using package {\sc
  GMoments}\footnote{\url{https://github.com/simoncasassus/GMoments}}. This
package fits the line profile in each line of sight with either one or
two Gaussians, and uses these model Gaussian line profiles to
calculate the velocity moments. For a single Gaussian with $I_v =
I_\circ \exp\left(-\frac{v-v^\circ}{\sigma_v}\right)$ the line
intensity is given by moment 0, or $I_{\rm Gauss} = \sqrt{2\pi}
I_\circ \sigma_v$, while the velocity centroid matches the Gaussian
centroid.  We note that the two-armed grand-design spiral that stands
out in the $^{12}$CO velocity-integrated intensity and velocity
dispersion, but that is absent in the line peak intensity. These
$^{12}$CO data do not appear to be affected by the underlying
continuum in the large crescent: continuum subtraction does not lead
to any local decrement in the tapered images, and $^{12}$CO(2-1)
emission is seen to extend out to radii of $\sim$1\farcs1. We note,
however, an arc-like decrement seen in the Gaussian amplitude in
Fig.\,\ref{fig:sgaussmoms}f, that is abut on the continuum so
coincident with the gap in between the two rings. Another particularly
interesting feature of the channel maps is the local $^{18}$CO(2-1)
peak coincident with the location of the candidate filament
(Fig.\,\ref{fig:sgaussmoms}n). This local peak is particularly strong
in the $^{18}$CO(2-1) channel maps at $v_{\rm lsr} \sim 8.5$km\,s$^{-1}$.


\begin{figure}
  \centering
  \includegraphics[width=\columnwidth,height=!]{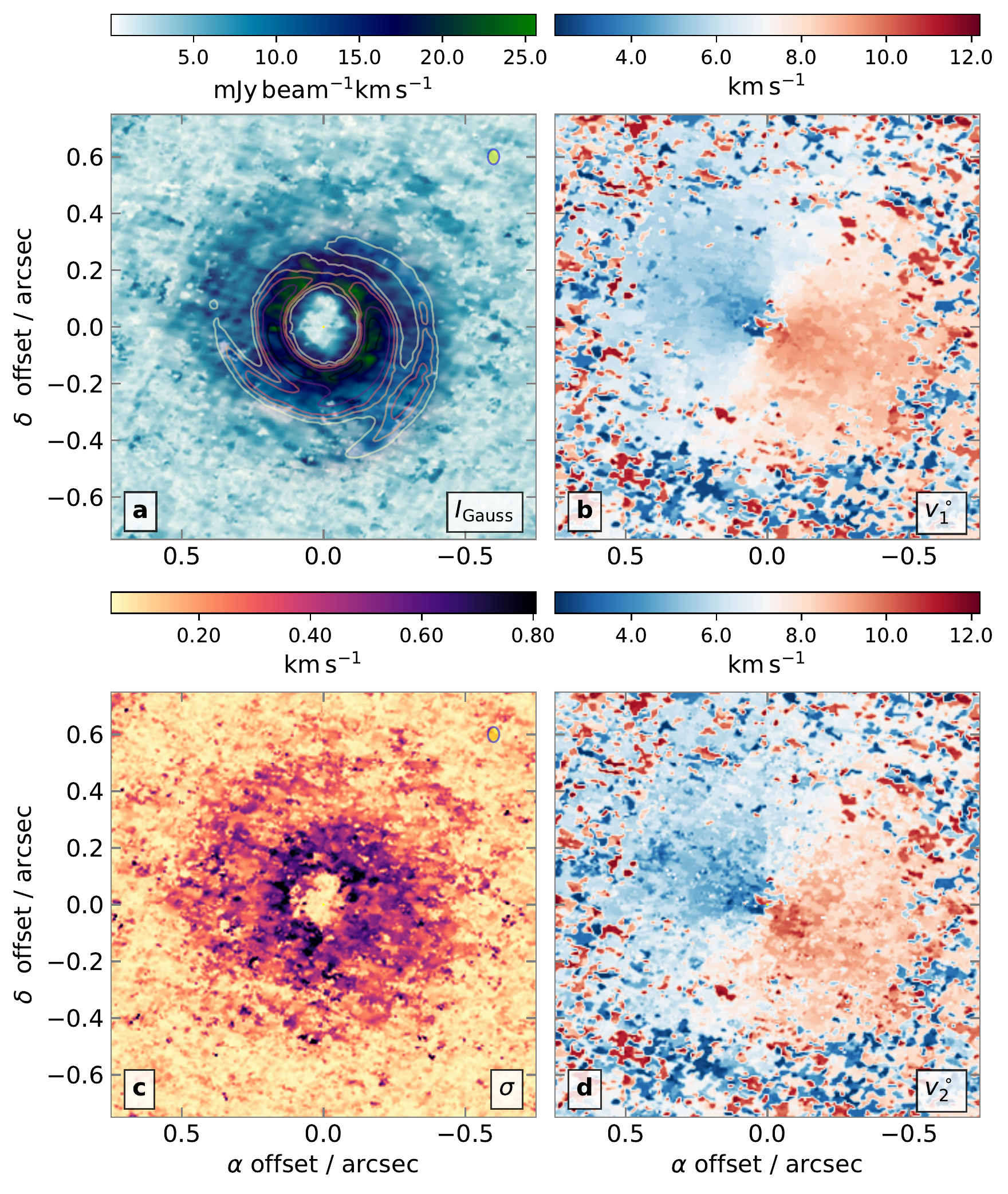}
  \caption{Moment maps in $^{12}$CO(2-1) extracted from the {\sc
      uvmem}-restored datacube using double-Gaussian fits in velocity,
    and comparison with the $H-$band polarised intensity.  {\bf a)}
    The 2-Gaussian velocity-integrated intensity in colour scale,
    compared with the $H-$band $Q_\phi$ image from
    Fig.\,\ref{fig:Qphi}, traced with the same contours (see Sec.\,\ref{sec:spirals}). {\bf b)} The
    velocity centroid of the brighter Gaussian. {\bf c}) The Gaussian
    dispersion. {\bf d)} The velocity centroid of the fainter
    Gaussian.  } \label{fig:12COuvmem}
\end{figure}

The top and bottom $^{12}$CO(2-1) layers are sufficiently separated in
this disc to trace each layer with a double-Gaussian fit, as
summarised in Fig.\,\ref{fig:12COuvmem} for the untapered {\sc uvmem}-restored
datacube. We assume that the brighter of the two
Gaussians traces the top side, that faces the observer.  The
velocity-integrated intensity is very similar to the single-Gaussian
case, as is the Gaussian velocity dispersion. Interestingly the disc
PA inferred from the velocity centroid of the brighter Gaussian,
$v^\circ_1$, shifts slightly to the North-West with increasing
distance from the star, as expected for the surface of a cone in which
the side nearest to the observer is to the South-East. However, the
velocity centroid $v^\circ_2$ corresponding to the fainter Gaussian,
shifts progressively to the South-East, indicating that it is indeed
tracing the bottom layer. This suggests that the extended disc and the
spiral modulation seen in dispersion is intrinsic to the top (or
bottom) layer, and is not the result of broadening due to a second
velocity component stemming from the bottom layer.



\section{Analysis} \label{sec:analysis}

\subsection{Continuum filament and disk orientation} \label{sec:analysis_continuum}


The 225\,GHz continuum image reported in
Fig.\,\ref{fig:HD135344B_cont} reveals interesting fine structure in
the gap that divides the inner and outer rings. A very fine filament
appears to join the two rings, at a radial separation of
$\sim$0\farcs468 (63.5\,au) and PA$\sim$ 216.1\,deg, as highlighted in
Fig.\,\ref{fig:HD135344B_cont}b. It is best seen in the {\tt tclean}
image with Briggs $r=0$, where the median intensity in a region
isolating the filament is 104\,$\mu$Jy\,beam$^{-1}$, and the 
noise is 24\,$\mu$Jy\,beam$^{-1}$. While this would appear as a
4$\sigma$ detection, there are other fine features in the same gap at
a similar intensity level, although these other features are smaller
and appear to sprout away from either the inner or outer ring (and may
also be real). These other features suggest that the detection of this
filament should be considered as a tentative result, whose
confirmation requires deeper observations. We note however in
Sect.~\ref{sec:spirals} that the filament is similar in shape and
pitch angle as a trailing spiral arm, and that it is almost coincident
with a twist observed in the near-IR spirals reported in
\citet{Muto2012ApJ...748L..22M} and
\citet{Stolker2016A&A...595A.113S}, which supports the idea that  the filament, if real,  may
 be tracing  gap-crossing planetary wakes.

\begin{figure}
  \centering
  \includegraphics[width=\columnwidth,height=!]{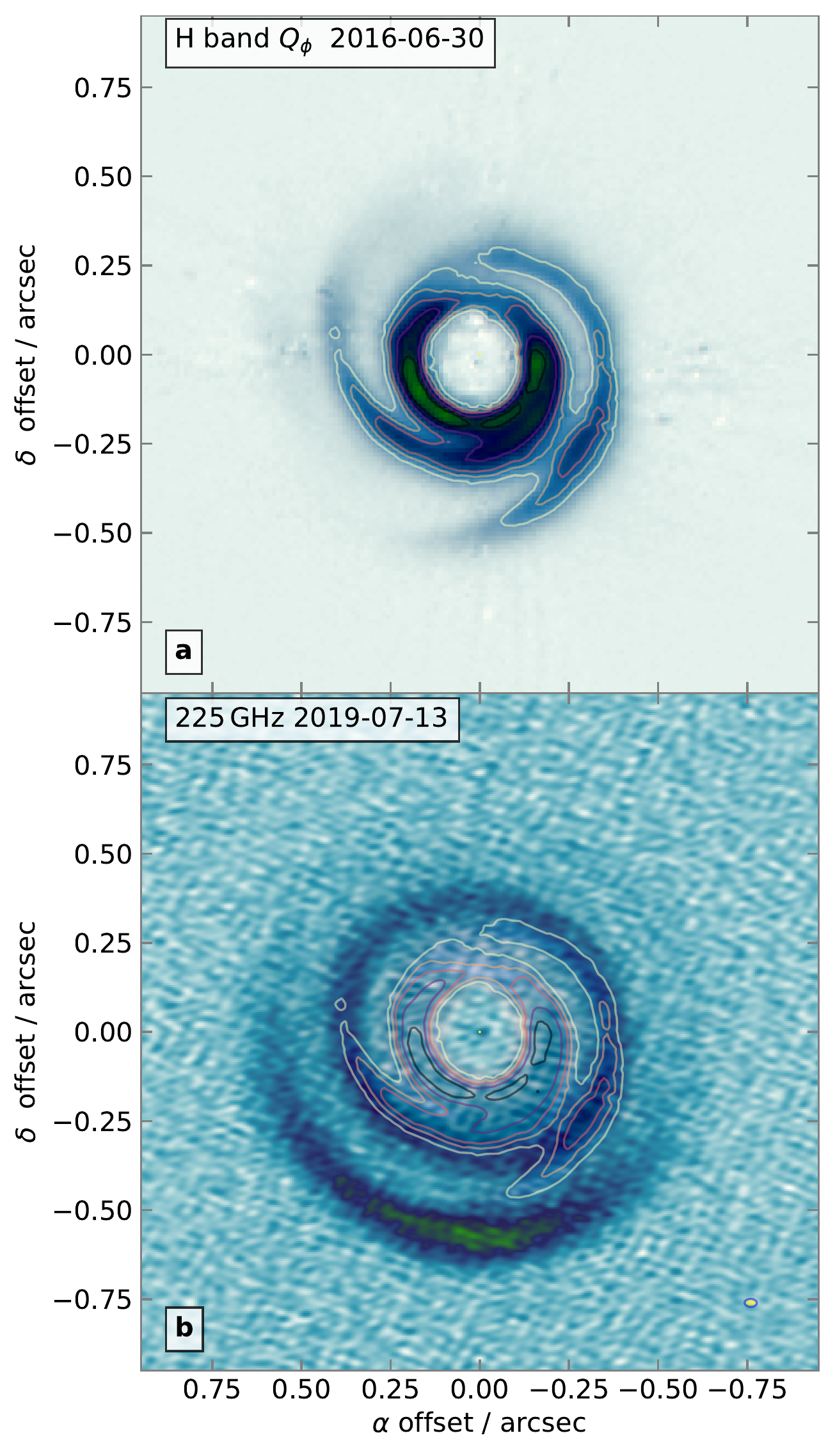}
  \caption{{\bf a:} IRDAP reduction of the H-band $Q_\phi$ image data
    from \citet[][]{Stolker2017ApJ...849..143S}, with contours taken
    at peak-intensity fractions of 0.2, 0.3, 0.4, 0.6, 0.8.{\bf b:}
    Overlay of the same contours as in a) on the Briggs $r=0$ image
    from Fig.\,\ref{fig:HD135344B_cont}b.} \label{fig:Qphi}
\end{figure}

Fig.\,\ref{fig:Qphi} compares the ALMA continuum image
of HD\,135344\,B with the polarised intensity image of the disc that was acquired
with VLT/SPHERE 
on 2016-06-30.
This $Q_{\phi}$ image \citep[see definitions in ][]{Avenhaus-2014b,
Garufi-2014} of HD\,135344\,B is the one obtained in the best seeing 
conditions \citep[average 0.5-$\mu$m seeing of 0\farcs37;][]{Stolker2017ApJ...849..143S},
and is therefore favoured throughout the rest of this work for comparison to our
ALMA data. We re-reduced this SPHERE/IRDIS dataset
with the {\sc irdap} pipeline \citep{VanHolstein2020A&A...633A..64V} to
produce the image shown in Fig.\,\ref{fig:Qphi}a.
 We see in
Fig.\,\ref{fig:Qphi}b that the contours that trace the $Q_\phi$ image
come very close to the (possible)  radio-mm  filament. 
While we leave a detailed comparison between the filament and the spiral
arms to Sect.~\ref{sec:spirals}, Fig.~\ref{fig:Qphi}b shows another 
interesting similarity between the radio-mm and IR in this source.
The faint continuum pedestal abut inside the inner ring appears to
surround the bright inner ring in polarised intensity. Such faint
pedestals are also seen in other systems, as, for example, in the
rings of DoAr\,44 and RXJ\,1633.9 \citep[called inflection points
  in][]{Cieza2021MNRAS.501.2934C} and in PDS\,70
\citep[][]{Isella2019ApJ...879L..25I}.


\begin{figure*}
  \centering
  \includegraphics[width=0.7\textwidth,height=!]{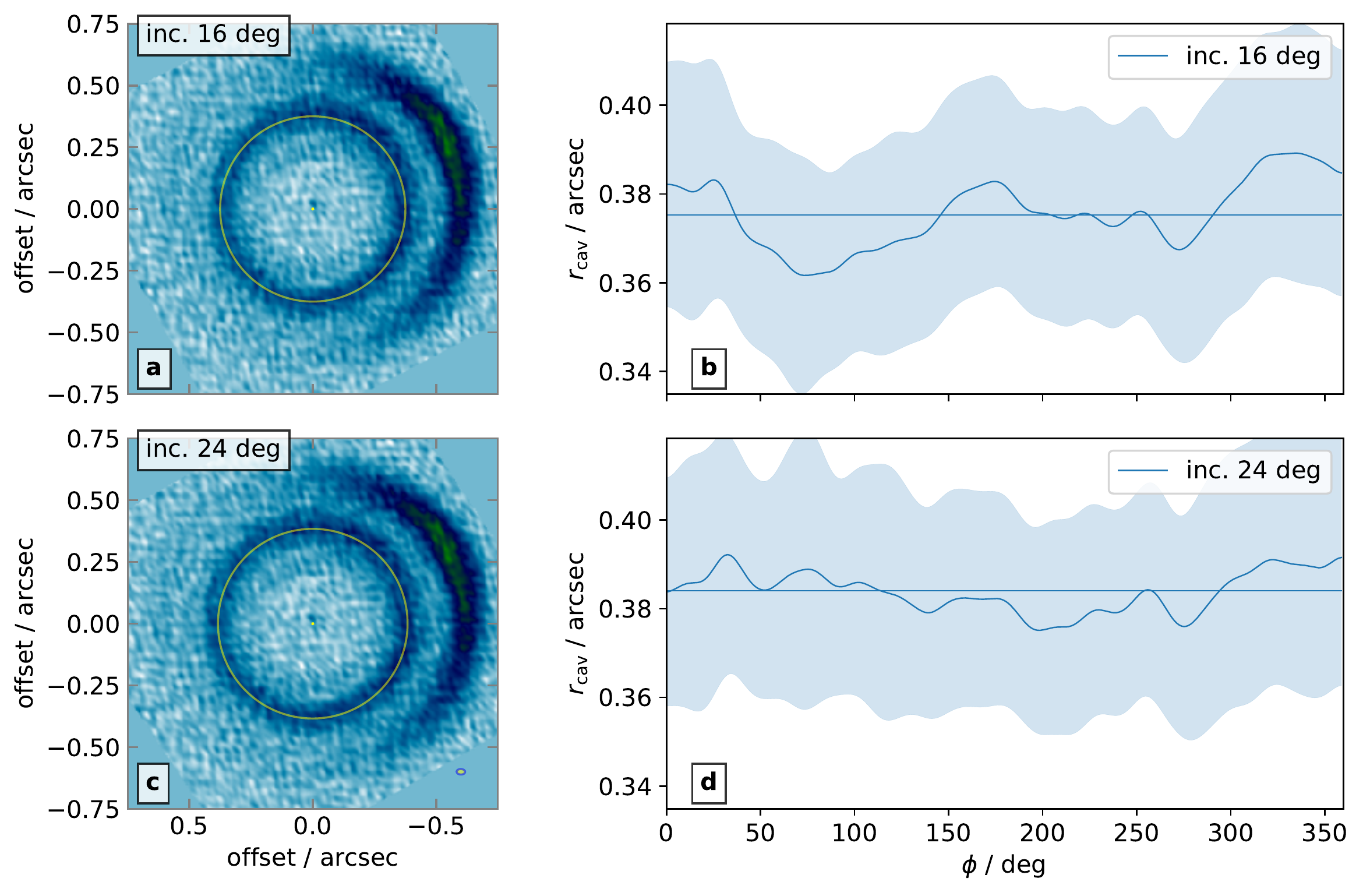}
  \caption{{\bf a:} face-on view of the Briggs $r=0$ image from
    Fig.\,\ref{fig:HD135344B_cont}b, assuming an inclination of
    16\,deg. {\bf b:} the azimuthal profile of the Gaussian centroid
    and dispersion in the inner ring, in a polar expansion of the
    image in a), after smoothing over two beam major axis (the total
    height of the shaded area corresponds to 1\,$\sigma$). {\bf c:}
    and {\bf d:} same as a) and b), for the optimal orientation of 24\,deg.} \label{fig:faceon_eccentric}
\end{figure*}

The continuum images can be used to infer disc orientation. We
minimised the scatter in the radial profile, extracted by averaging in
azimuth over the radial range $[0\farcs25,0\farcs45]$ (so enclosing
only the inner ring). The procedure involves 4 free-parameters: the
disc position angle (PA), inclination $i$, and the origin for the
polar expansions, which is offset relative to the origin of
coordinates in the images by $\Delta \alpha$ in right-ascension and
$\Delta \delta$ in declination. We used the {\sc MPolarMaps} package,
which is described in appendix. The posterior distributions were
sampled with the {\sc emcee} package
\citep[][]{emcee2013PASP..125..306F}, using flat priors. The disc
orientation resulting from the tclean image with $r=0$ Briggs weights
is PA$=241.6^{+6.9}_{-6.0}$\,deg, $i=23.80^{+2.5}_{-2.6}$\,deg,
$\Delta \alpha = 2^{+3}_{-4}$\,mas, $\Delta \delta =
12^{+4}_{-4}$\,mas. The origin of the polar expansion is plotted with
a red marker in Fig.\,\ref{fig:HD135344B_cont}. We note that the disc
PA is consistent with that used by
\citet[][]{Cazzoletti2018A&A...619A.161C}, considering that here we
point PA at the position of the ascending node, but the disc
inclination inferred in this work is significantly higher. The face-on
views in Fig.\,\ref{fig:faceon_eccentric} show that the present
inclination results in a circular inner ring, albeit offset from the
central emission. The orientation that corresponds to the most axially
symmetric ring results in a scatter for the azimuthal profile of the
ring radius of $\sigma(r_{\rm cav} ) = 5$\,mas about a median of
$0\farcs38$. However, a lower inclination of $i=16$\,deg results in a
pronounced eccentricity in the deprojected (face-on) views, which is
more conspicuous in the polar expansions, with an azimuthal scatter
$\sigma(r_{\rm cav} ) = 8$\,mas, about a median of $0\farcs37$.


The high inclination resulting from the present analysis of the
continuum may reflect that the inner ring is intrinsically
eccentric. As discussed below (Sec.\,\ref{sec:rotcurve}), an
inclination close to $i=16$\,deg is required to bring the dynamical
mass of the star in agreement with photospheric measurements. The
difference with the inferred inclination of
$i=23.80^{+2.5}_{-2.6}$\,deg based on the inner dust ring suggests that
the inclination difference, of $\sim 7.8 \pm 2.5$, is due to intrinsic
ring eccentricity, which corresponds to $e = 0.14\pm0.04$. This value
is comparable to the measurement of $e\sim 0.1$ in MWC\,758 by
\citet[][]{Dong2018ApJ...860..124D}.

%
%
%





The difference in the continuum inclination derived here with the 
lower values reported in \citet[][]{Cazzoletti2018A&A...619A.161C},
might be due to their use of a parametric model that results in large
visibility residuals. In their noisiest data, those from ALMA Band\,3,
the residuals are adequately thermal and the inclination is
$i=17.7^{+3.4}_{-4.7}$\,deg, so consistent with our value of $i=23.80^{+2.5}_{-2.6}$\,deg given the
error bars. However, their value with the smallest errors, of
$i=9.8\pm0.1$\,deg in Band\,7, results in non-thermal residuals
suggesting that the parametric model does not provide a good fit and
that these uncertainties are thus artificially low. Another possible
source of discrepancy is that
\citet[][]{Cazzoletti2018A&A...619A.161C} kept the disk position angle
fixed in their optimisation, thereby reducing the uncertainties on
inclination. Whichever the source of the bias, an inclination of
$i=9.8\pm0.1$\,deg would yield much too high dynamical stellar masses
(see Sec.\,\ref{sec:rotcurve}).

%
%

\subsection{Spiral arms} \label{sec:spirals}

\begin{figure*}
  \centering
  \includegraphics[width=0.33\textwidth,height=!]{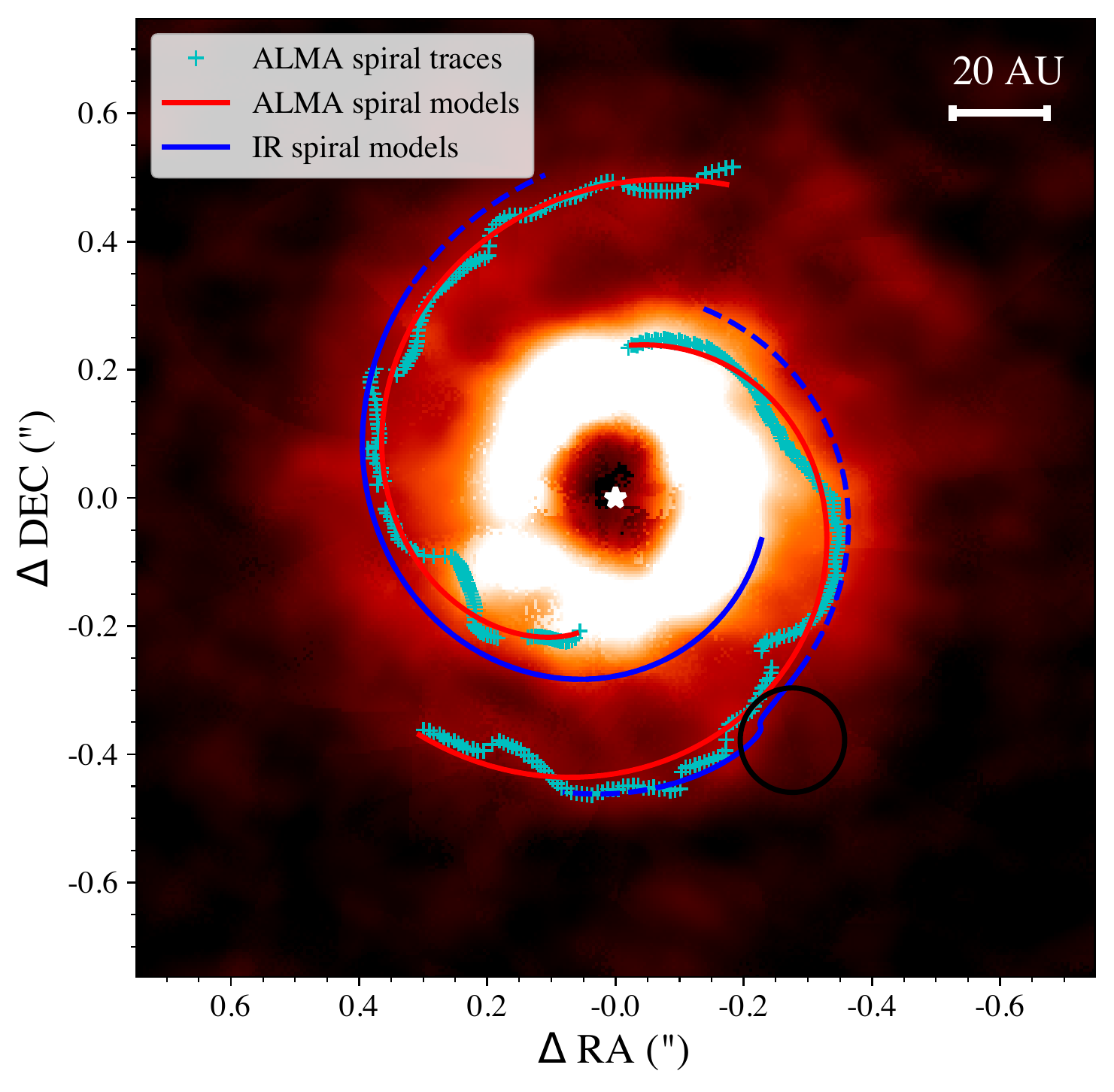}
  \includegraphics[width=0.33\textwidth,height=!]{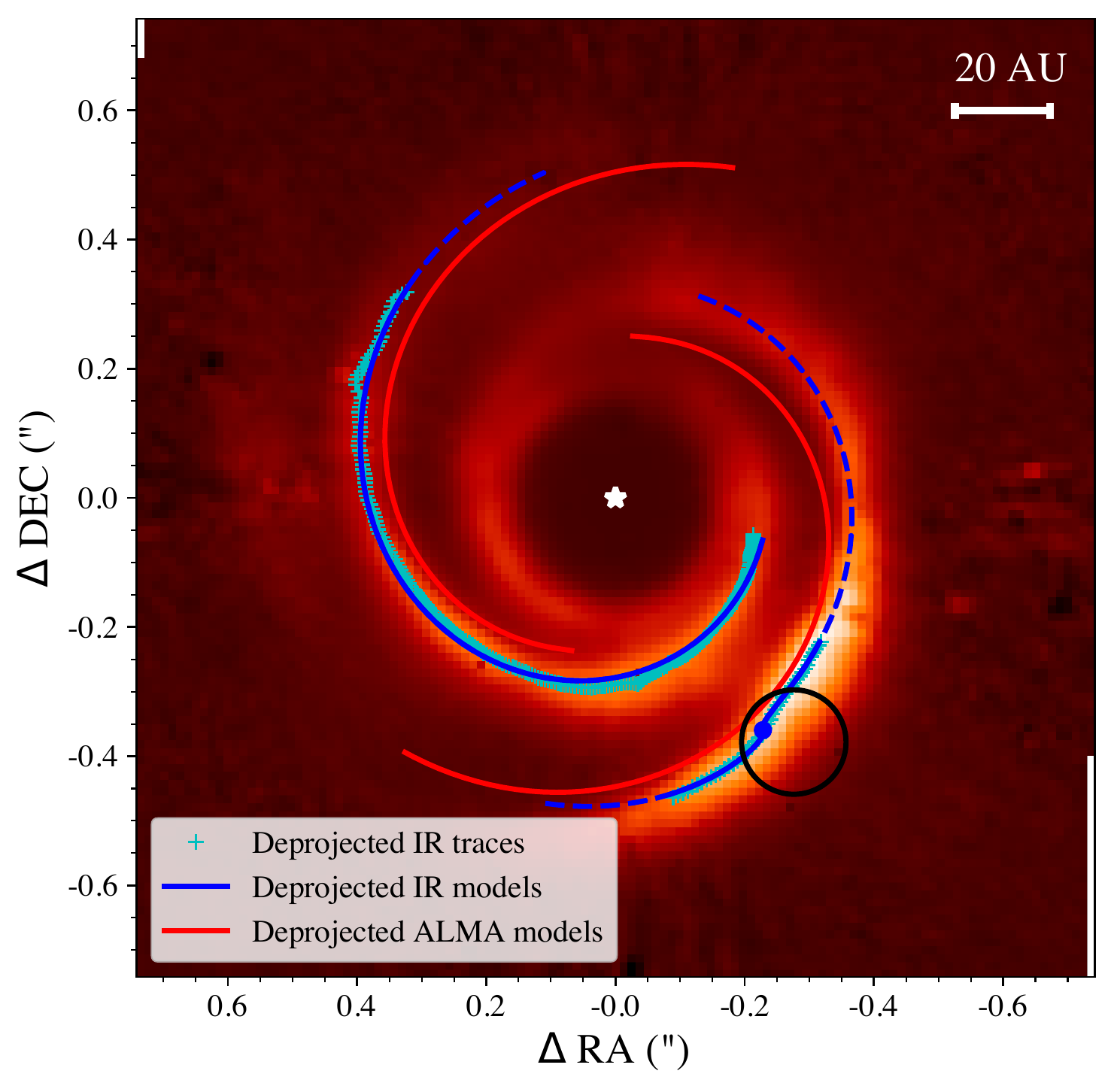}
  \includegraphics[width=0.33\textwidth,height=!]{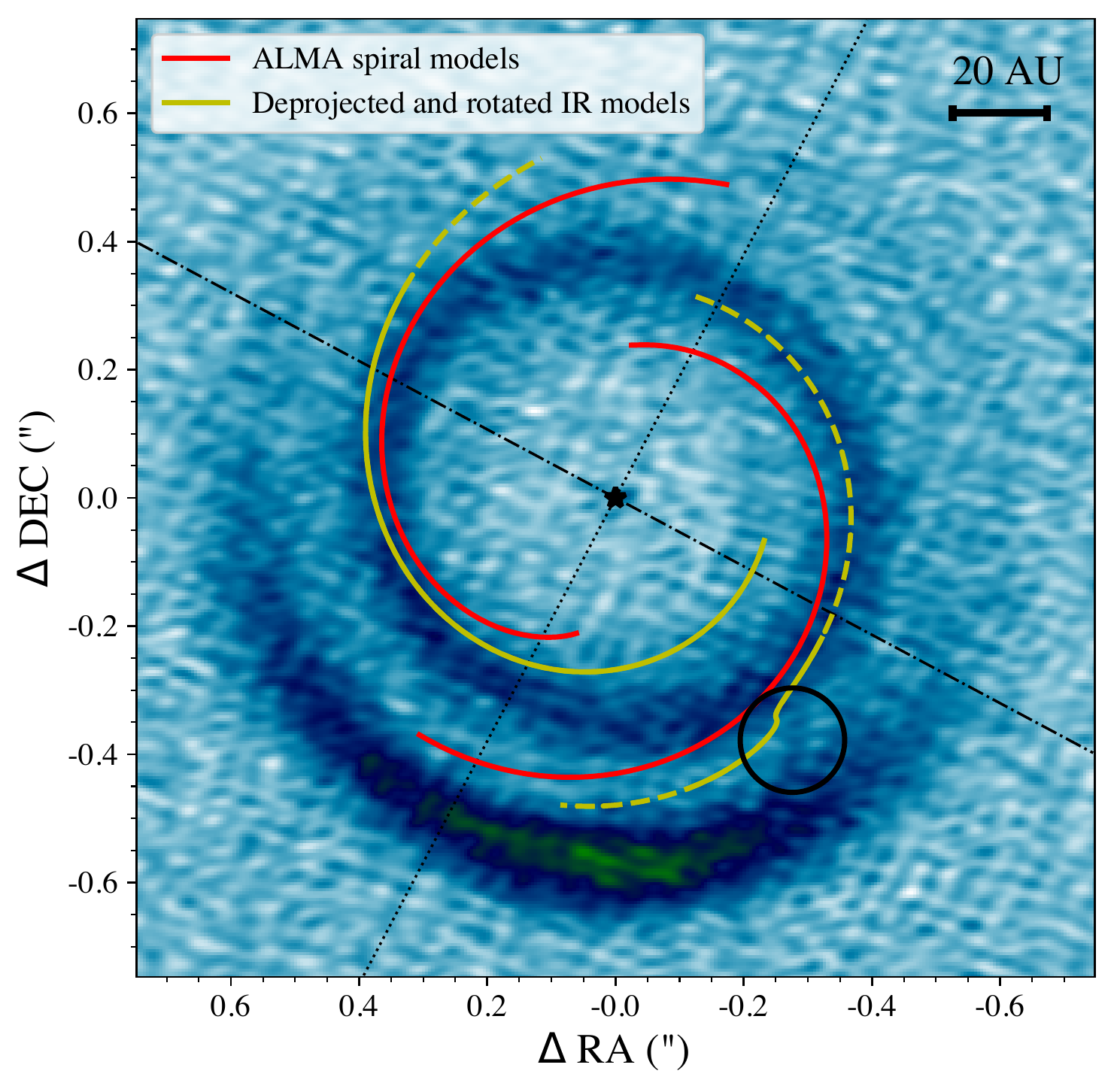}
  \caption{Spiral traces (cyan crosses) identified in the velocity-integrated
    $^{12}$CO(2-1) intensity map $I_{\rm Gauss}$($^{12}$CO(2-1)) and the near-IR scattered-light image, and corresponding best-fit spiral models compared
    to $I_{\rm Gauss}$($^{12}$CO(2-1)) (left), the deprojected near-IR
    scattered-light image (middle) and the sub-mm continuum image
    (right). The location of the gap-crossing sub-mm continuum
    filament is circled in all panels. All images are shown in linear
    scale, with cuts corresponding to the 5th and 95th percentiles of
    pixel intensities for the $^{12}$CO(2-1) moment 0 map, and min/max
    cuts for the IR and continuum images.  The IR spiral models are
    shown projected (i.e. as observed) in the left panel, and
    deprojected onto the plane of the sky (i.e. after flaring and
    inclination correction) in the middle panel.  The best-fit
    location of the planet for the fit to Eq.~\ref{Eq:Rafikov2002} is
    provided with a blue dot in the middle panel. In the sub-mm
    continuum image, the deprojected model for the south-west IR
    spiral is rotated by 3.7\,deg in the prograde direction, and
    subsequently projected onto the disc mid-plane (i.e.~considering
    disc inclination, but not surface flaring). The semi-major and
    semi-minor axes are shown with dotted-dashed and dotted lines
    respectively.  } \label{fig:spiral_fit}
\end{figure*}

Both the native and tapered versions of the velocity-integrated
$^{12}$CO(2-1) intensity map $I_{\rm Gauss}$($^{12}$CO(2-1)) and its
dispersion $\sigma_v$($^{12}$CO(2-1)) reveal a double-armed spiral
pattern (Figs.\,\ref{fig:sgaussmoms} and \ref{fig:12COuvmem}),
reminiscent of the morphology of the near-IR spirals reported in
\citet{Muto2012ApJ...748L..22M, Stolker2016A&A...595A.113S}.  In order
to facilitate the interpretation of these spirals, we identified their
trace in both $I_{\rm Gauss}$($^{12}$CO(2-1)) and the $H-$band
$Q_\phi$ image, fitted these traces to different spiral equations, and
compared the best-fit models with $I_{\rm Gauss}$($^{12}$CO(2-1)), the
$H-$band image (hereafter the IR image) and the sub-mm continuum image
(Fig.\,\ref{fig:spiral_fit}).

We first identified local radial maxima in 1-deg wide slices in the $uv$-tapered $I_{\rm Gauss}$($^{12}$CO(2-1)) (where the spirals 
are the most conspicuous), after subtraction of the median radial intensity profile. These median intensities 
are integrated azimuthally over concentric 1-FWHM wide ellipses for all pixels, considering the mid-plane
orientation of the disc inferred from the CO rotation curves (Sect.~\ref{sec:lines}; $i \sim 16$\,deg 
and PA$\sim$241.9\,deg). 
In each azimuthal slice, the vicinity of each radial maximum identified is then fitted to a 1D Gaussian
profile using {\sc scipy}'s {\tt curve\_fit} routine. For each position angle, the radial separation and 
associated uncertainty on the spiral trace are set as the centroid of the Gaussian, and the maximum
between the uncertainty on the centroid and half the beam size of our observations, respectively.
The traces inferred with this procedure are shown as cyan crosses in Fig.\,\ref{fig:spiral_fit}a. We then fitted 
these traces to the equation of a general Archimedean spiral ($r = a + b\theta^{n}$), and found 
the best-fit parameters using  {\sc scipy}'s Nelder-Mead minimisation algorithm.
Since the objective of this fit is the comparison of the CO spirals with the IR spirals, we did not deproject 
the disc, hence the choice for a general Archimedean spiral equation for the fit.

We then proceeded to a similar analysis in IR image from
Sec.\,\ref{sec:analysis_continuum} (Fig.\,\ref{fig:Qphi}).  The
southwest spiral shows a sudden discontinuity in pitch angle,
identified as 'kink' in \citet{Muto2012ApJ...748L..22M} and
\citet{Stolker2016A&A...595A.113S}. They attributed the kink to the
location of a planet driving this spiral arm, which they labelled
`S1'. Therefore we adopted a different equation for the fit of this
specific spiral trace. Instead of a general Archimedean spiral, we
considered the expected shape from a spiral density wave excited by an
embedded planet in the linear or weakly non-linear regime
\citep{Rafikov2002ApJ...569..997R}:
\begin{align}
\label{Eq:Rafikov2002}
\theta(r) &= \theta_p + \frac{{\rm sign}(r-r_p)}{h_p} \nonumber \\ &\times
\Bigg\{ \left( \frac{r}{r_p} \right)^{1+\beta} \bigg[
  \frac{1}{1+\beta} - \frac{1}{1-\alpha +\beta} \left( \frac{r}{r_p}
  \right)^{-\alpha} \bigg] \nonumber \\ & - \left( \frac{1}{1+\beta} -
\frac{1}{1-\alpha +\beta} \right) \Bigg\}
\end{align}
where ($\theta_p$, $r_p$) are the polar coordinates of the planet location, $h_p$ is the disc aspect ratio at the planet's location, 
and $\alpha$ and $\beta$ are the exponents of power laws for the angular frequency of the disc ($\Omega \propto r^{-\alpha}$) 
and the sound speed ($c_s \propto r^{-\beta}$), respectively.
The five parameters of this equation are known to be highly degenerate \citep[e.g.][]{Muto2012ApJ...748L..22M,Christiaens2014ApJ...785L..12C}.
Furthermore, the linear or weakly non-linear approximation is only valid in the vicinity of the planet \citep[e.g.][]{Zhu2015},
with an increasingly larger pitch angle (with respect to the linear approximation) the further the separation with the planet location.
Given these considerations, we restricted the spiral trace to only a section subtending 45\,deg around the twist. 
Moreover, we fixed $\alpha$ to 1.5 (Keplerian rotation) and $\beta$ to 0.207, the average flaring found 
in \citet{Avenhaus2018ApJ...863...44A} based on their sample of $H$-band polarised intensity images of protoplanetary discs, and left $\theta_p$, $r_p$ and $h_p$ as 
free parameters.
Since Eq.~\ref{Eq:Rafikov2002} assumes a face-on view of the disc, we deprojected the near-IR image using {\sc diskmap}\footnote{\url{https://github.com/tomasstolker/diskmap}}\citep[][]{Stolker2016A&A...595A.113S},
adopting $\beta = 0.207$ and an initial estimate of 0.1 for $h_p$.
We also scaled the image by $r^2$ to compensate for stellocentric flux dilution.
We then proceeded with finding the optimal values of $\theta_p$, $r_p$ and $h_p$ with {\sc scipy}'s Nelder-Mead minimisation algorithm, iteratively updating the value of $h_p$ used for the disc deprojection.
The values of $h_p$ used for deprojection and found by the fit of the spiral traces converged to within 0.1\% relative values within 2 iterations.
The final values we inferred are a PA of $212.4\pm0.7$\,deg and $r = 425.9 \pm 1.2$ mas ($57.8 \pm 0.2$ au) for the planet's location, and a disc aspect ratio $h_p = 0.017 \pm 0.001$ at that location,
at the epoch of the IRDIS observations.
The uncertainties on each parameter were found by bootstrapping (1000 bootstraps).
We tested different values of $\beta$ ranging from 0 to 0.25 and found consistent values of $\theta_p$, $r_p$ and $h_p$.
Fig.\,\ref{fig:spiral_fit}b shows the best-fit model onto the deprojected disc image, with intensities scaled by $r^2$.
We notice a tentative radial shift between the CO models and IR spirals, which is likely due to the 
different emitting and scattering surfaces for $^{12}$CO(2-1) and sub-$\mu$m size dust grains, respectively \citep[e.g.][]{Pinte2018A&A...609A..47P,Avenhaus2018ApJ...863...44A}.

The best-fit disc aspect ratio for the scattering surface, $h_p =
0.017\pm0.001$ appears significantly smaller than the expected value
based on the $^{12}$CO(2-1) brightness temperature ($h_g \sim 0.1$,
see Sec.\,\ref{sec:lines} and Fig.~\ref{fig:props}d). One possibility
to account for both the very low inferred value of $h_p$ for the IR
spiral fit and the small apparent radial shift between the IR twist
and the sub-mm continuum filament is that the outer spiral arm (with
respect to the twist) is more curled-in towards the star at the
$H$-band scattering surface than the inner spiral \citep[as e.g in the
  3D simulations of][]{Zhu2015}.  This may result from a geometrically
thick and vertically non-isothermal disk (i.e.~non-constant sound
speed at a given cylindrical radius). Alternatively, if the spiral
structure traces surface waves, it may also travel slower than sound
speed, hence inducing more tightly wound spirals than predicted by the
spiral density wave theory. Since the value of $h_p$ inferred in the
fit depends on the apparent radial amplitude of the twist, a more
curled in outer spiral would artificially reduce the inferred value of
$h_p$. Furthermore, this curling-in would also explain the radial
shift between the outer part of S1 and the azimuthal asymmetry in the
outer sub-mm continuum ring. The latter appears to be located in the
continuity of the filament, and it may thus be tracing dust following
the gas density enhancement in the planetary wake, which may be
possible for low Stokes number \citep[see e.g.][]{Veronesi2019}. In
this scenario, the spiral wake models of Eq.~\ref{Eq:Rafikov2002} (as
used in Fig.~\ref{fig:spiral_fit}) would suggest a mid-plane
temperature of the order of $\sim$25~K ($h_p \sim 0.07$) to follow the
pitch angle of the candidate filament and join the outer ring. It is
worth noting however that the radial dependence of the sound speed
(and hence the temperature profile) is assumed to be a power law in
the derivation of Eq.~\ref{Eq:Rafikov2002}, with a constant power
index equal to -$\beta$.  Therefore, if the sound speed (or more
broadly the wave propagation speed) follows a more complex radial
profile, the spiral morphology would deviate accordingly and would
possibly follow more closely the shape of the outer ring asymmetry.

Finally, we compared all spiral models to our sub-mm continuum image
of the disc (Fig.\,\ref{fig:spiral_fit}c).  Taking into account
Keplerian rotation around a star with mass $M_\star =
1.67\,^{+0.18}_{-0.16} M_\odot$
\citep[][]{Wichittanakom2020MNRAS.493..234W}, and for a distance of
$135.7\pm1.4$\,pc \citep[][]{Gaia2018A&A...616A...1G}, the difference
of 3.04\,yrs 
\citet{Muto2012ApJ...748L..22M} between the epochs of the IRDIS and
ALMA data corresponds to a prograde rotation of
$\sim$3.7\,deg.
amount in the right panel of Fig.~\ref{fig:spiral_fit}.  More
precisely, we used the deprojected IR model (shown in the middle panel
of Fig.~\ref{fig:spiral_fit}) for the rotation and subsequent
re-projection onto the disc mid-plane (i.e. considering the
inclination of the disc, but not the original flaring) -- see left
panel of Fig.~\ref{fig:spiral_fit} for the relative locations of the
non-deprojected spiral model with respect to the filament.  This
angular shift of 3.7\,deg nicely aligns the PA of the tentative
filament ($\sim$216.1\,deg; Sect.~\ref{sec:analysis_continuum}) and
that of the IR spiral arm twist ($216.1\pm0.7$\,deg after rotation).
However, we measure a radial shift $\gtrsim42.1\pm 1.2$\,mas between
the centre of the filament ($\sim$468\,mas) and the planet location
inferred from our fit to the IR twist, even after deprojection.  The
ALMA pointing uncertainty appears insufficient to account for this
radial shift, as it is $\lesssim$15 mas, e.g. if the sub-mm continuum
signal near the center of the cavity is tracing the star (see inset of
Fig.~\ref{fig:HD135344B_cont} a).  As mentioned above, the observed
radial shift may be consistent with the expected curling of the spiral
wave towards the star \citep[e.g.][]{Zhu2015} as  the
IR scattering surface would  be located at a shorter deprojected radius
than the bulk of the density wave (in the mid-plane and likely consistent with the
locus of the continuum filament).  It is worth noting that given the likely
eccentric geometry of the inner ring
(Sec.~\ref{sec:analysis_continuum}), the planet may also be on a
slightly eccentric orbit, which may also partially contribute to this
offset.

\subsection{Line diagnostics of physical conditions}  \label{sec:lines}

The uniform slab approximation goes a long way in observational
astronomy, as it is a simple means to extract physical conditions
in a given line-of-sight. We use this approximation to estimate
physical conditions in the gas using the CO isotopologue rotational
lines. In local-thermodynamic-equilibrium (LTE) the emergent intensity
from ground-state rotational lines depends on the column of the
emitting specie, on the uniform-slab temperature $T_b$, and on the
line-of-sight turbulent broadening $v_\mathrm{turb}$. We write the gas
temperature with an under-script `b' to remind that, in the case of CO,
this temperature is close to  the brightness temperature of the
optically thick $^{12}$CO. Given fractional abundances, the emitting
column can be converted into a total gas surface density,
$\Sigma_g$. We have developed a tool to fit multi-isotopologue data
with these free  parameters, which we call {\sc Slab.Line}. Related
approaches have also been considered by others
\citep[][]{Teague2016A&A...592A..49T,Flaherty2020ApJ...895..109F,
  Garg2020arXiv201015310G, YenGu2020ApJ...905...89Y}. The model line
profile for a given line of sight $\vec{x}$ is
\begin{equation}
I^m_\nu(\vec{x}) = B_\nu(T_b(\vec{x})) \left[  1 - \exp\left( -\tau_\nu(\vec{x}) \right)  \right],
\end{equation}
as a function of frequency $\nu$, with
\begin{equation}
\tau_\nu(\vec{x})= \kappa_L(\vec{x}) \Sigma_g(\vec{x}) f_{\rm mol} \Phi_\nu(\vec{x}). \label{eq:tau_L}
\end{equation}
The line opacity $\kappa_L(\vec{x})$ is approximated in LTE, so  for a rotational transition $J_2 \rightarrow J_1$ 
\begin{equation}
  \kappa_L(\vec{x}) = \frac{ h  \nu_\circ}{ 4 \pi m_{ {\rm H}_2} }
  \frac{g_{J_1} e^{-\frac{E_{J_1}}{k T_b(\vec{x})}}}{Z(\vec{x})} B_{12}
  \left[  1 - e^{ - \frac{ h \nu_\circ }{ k T_b(\vec{x})}} \right]. 
\end{equation}
We use the LAMDA molecular database
\citep[][]{LAMDA2005A&A...432..369S}, and evaluate the partition
function $Z = \sum_{J=0}^{J_{\rm max}} g_{J} e^{-\frac{ E_{J}}{k
    T_b}}$ by summing over all tabulated rotational energy levels
$E_{J}$ (so for $^{12}$CO this corresponds to $J_{\rm max} = 40 $).
In Eq.\,\ref{eq:tau_L}, $f_{\rm mol}$ is the abundance by number of
the emitting molecule relative to H$_2$. In this case for the CO
isotopologues we set $f_{^{13}{\rm CO}} = \frac{1}{76} f_{^{12}{\rm
    CO}} $ \citep[from][]{Casassus2005A&A...441..181C,
  Stahl2008A&A...477..865S}, $f_{{\rm C}^{18}{\rm O}} =
\frac{1}{500}f_{^{12}{\rm CO}} $ \citep{Wilson1994}, with
$f_{^{12}{\rm CO}} = 10^{-4}$. The line profile is simply a thermal
Gaussian broadened by turbulence, with dispersion $v_{\rm  turb}(\vec{x})$ and velocity centroid $v_\circ(\vec{x})$:
\begin{equation}
\Phi_{\nu}(\vec{x}) = \frac{1}{\sqrt{2\pi}\sigma_{\nu_\circ}} \exp\left(  - \frac{ \left( \nu - \nu_{\circ}\right)^2}{2 \sigma_{\nu_\circ}^2}  \right),
\end{equation}
with $\nu_\circ = \frac{E_{J_2} - E_{J_1}}{h} \left( 1 + \frac{v_{\circ}}{c} \right)$ and 
\begin{equation}
  \sigma_{\nu_\circ}(\vec{x}) = \frac{\nu_\circ}{c} \sqrt{ \frac{k T_b(\vec{x})}{m_{\rm mol}} + v_{\rm turb}^2(\vec{x})},
\end{equation}
and where $m_{\rm mol}$  is the molecular mass.

The free parameters $v_\circ$, $\Sigma_g$, $T_b$ and $v_{\rm turb}$ were constrained in each line of sight $\vec{x}_l$ independently with a least-squares fit to the observed spectra in the three isotopologue transitions:
\begin{equation}
  \chi^2(\vec{x}_l) = \sum_{i=1}^{3}  \frac{1}{I^2_{{\rm rms, i}}(\vec{x}_l)} \sum_{\nu_k} \left( I_{\nu_k}(\vec{x}_l) - I_{\nu_k}(\vec{x}_l)^m    \right)^2, \label{eq:slablinechi2}
\end{equation}
where the sum in frequencies $\left\{\nu_k\right\}$ runs over all
available spectral channels. The noise in each line of sight,
$I_{{\rm rms}, i}(\vec{x}_l)$, is taken as the rms dispersion in the
observed spectra $\pm1\,$km\,s$^{-1}$ away from the peak of the
line. The optimisation for each line of sight was carried out in the
logarithm of the positive-definite parameters, i.e. the full set of
parameters is $(\log_{10}(\Sigma_g), \log_{10}(T_b), v_\circ,
\log_{10}(v_{\rm turb}))$. An application of the conjugate-gradient
method, as implemented in  {\sc scipy.optimize}, yielded a first
approximation to adequate sets of parameters. We then sampled
parameter space with the {\tt emcee} package
\citep[][]{emcee2013PASP..125..306F}, which is a Markov chain Monte
Carlo ensemble sampler \citep{MCMC2010CAMCS...5...65G}, using flat
priors.

\begin{figure*}
  \centering
  \includegraphics[width=0.7\textwidth,height=!]{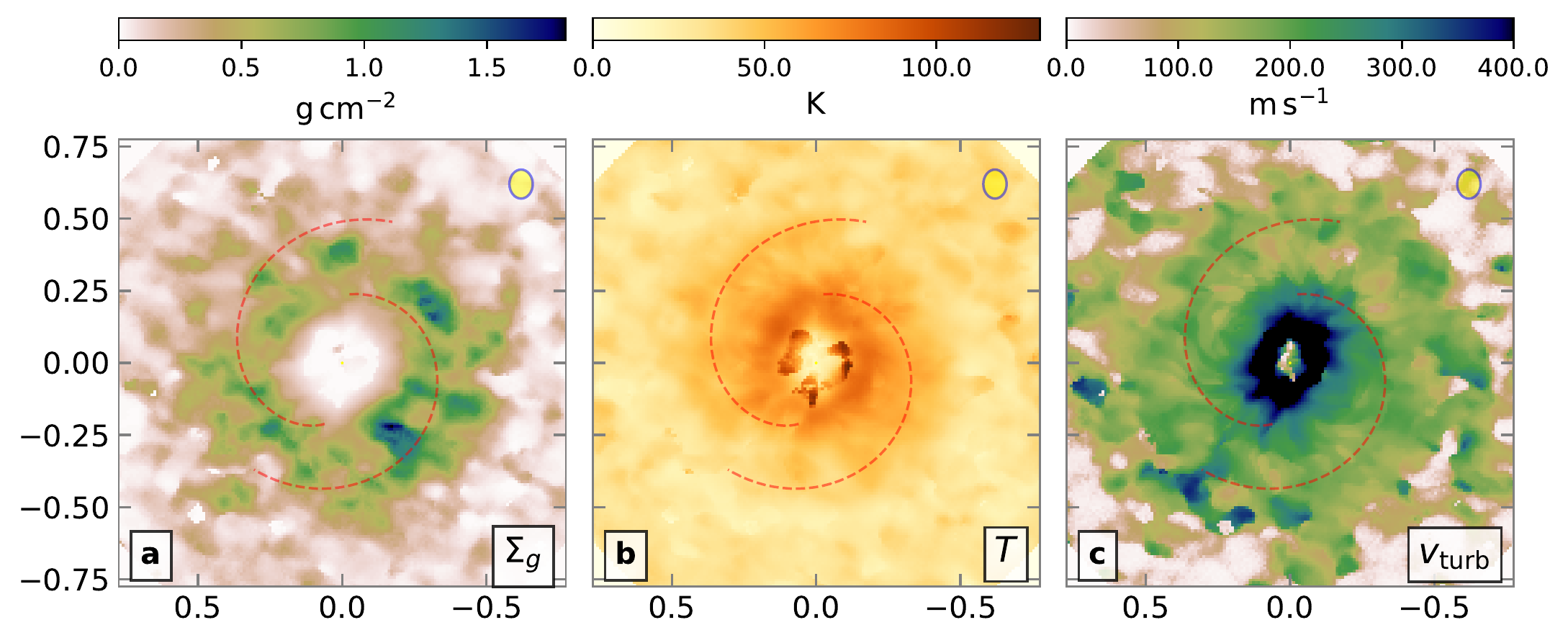}
  \includegraphics[width=0.7\textwidth,height=!]{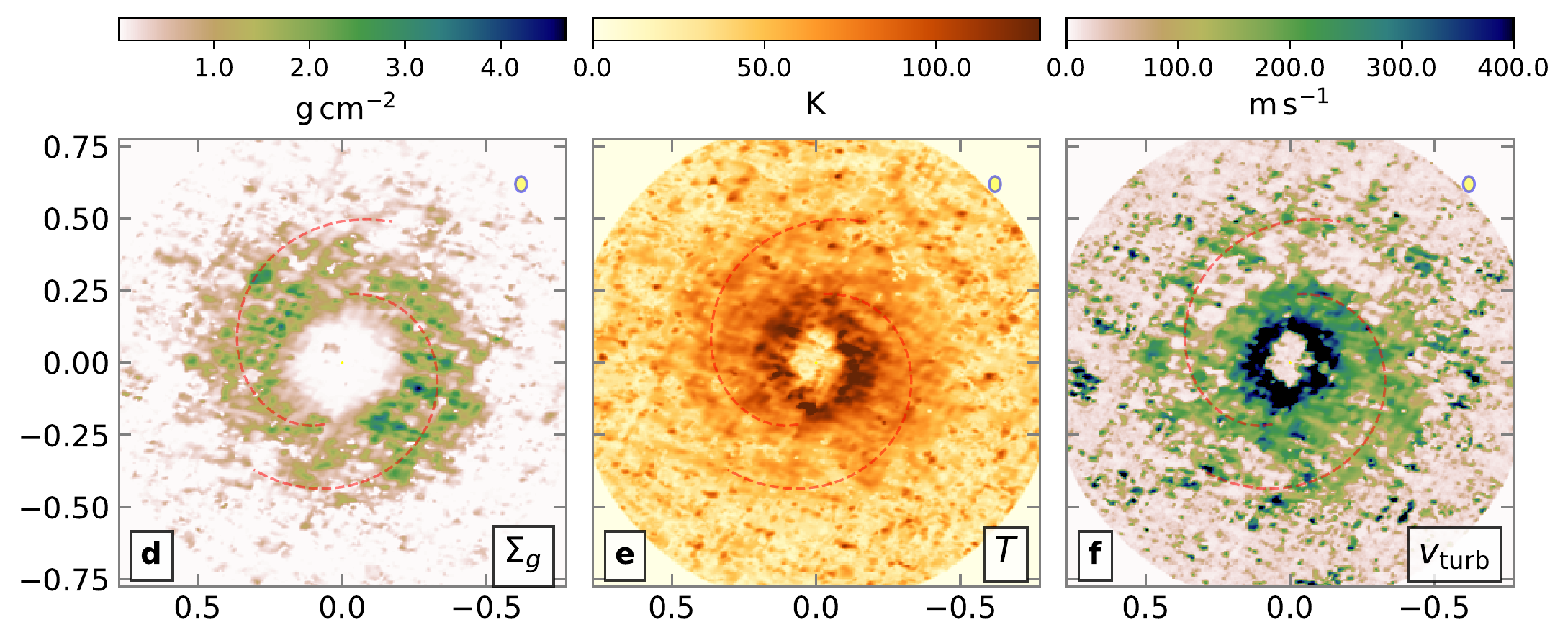}
  \caption{Example application of {\sc Slab.Line} to extract physical
    conditions in HD\,135344B using the $^{12}$CO(2-1), $^{13}$CO(2-1)
    and C$^{18}$O(2-1) lines.{\bf a,b,c):} measurements inferred from
    the $uv$-tapered datacubes.  {\bf a)}: total gas surface density
    $\Sigma_g$, {\bf b)}: gas temperature $T_b$, {\bf c)}: turbulent
    line broadening $v_{\rm turb}$. {\bf d,e,f):} same as a,b,c but
    for the original datacubes without $uv$-tapering.  The dashed red
    lines corresponds to the traces of the CO spirals obtained in
    Sec.\,\ref{sec:spirals}.} \label{fig:linefit_HD135344B}
\end{figure*}

An application of {\sc Slab.Line} to HD\,135344B is shown in
Fig.\,\ref{fig:linefit_HD135344B}, where we report the fields for
$\Sigma_g(\vec{x})$, $T_b(\vec{x})$ and $v_{\rm turb}(\vec{x})$
inferred from both the natural-weights datacubes and the $uv-$tapered
datacubes.   These fits assumed fixed isotopologue
abundances, but we reach thermal residuals nonetheless. We refer to
Appendix\,\ref{sec:slab.line} for a discussion of goodness of fit,
correlation analysis, and example lines of sights.

In Fig.\,\ref{fig:linefit_HD135344B} it is particularly interesting to
note the similarity between $v_{\rm turb}$ and the line velocity
dispersion map $\sigma(\vec{x})$ in Figs.\,\ref{fig:sgaussmoms} and
\ref{fig:12COuvmem}, as both follow the spiral pattern discussed in
Sec.\,\ref{sec:spirals}, but $\Sigma_g$ and $T_b$ do not. The larger
scale spiral is best traced in the $uv$-tapered version of $v_{\rm
  turb}(\vec{x})$, while the root of the spirals is seen in the native
version (with no $uv$-taper).

Another interesting feature of the line diagnostics in
Fig.\,\ref{fig:linefit_HD135344B} is the absence of a local peak
neither $\Sigma_g$ nor $T$ at the position of the filament. This is
surprising because this position coincides with the maximum line
intensity in $^{18}$CO(2-1) (see Sec.\,\ref{sec:obs}). A possible
interpretation is that the mid-plane near the candidate is hotter than
the surface sampled in $^{12}$CO(2-1).


%
%
%
%

\begin{figure}
  \centering
  \includegraphics[width=\columnwidth,height=!]{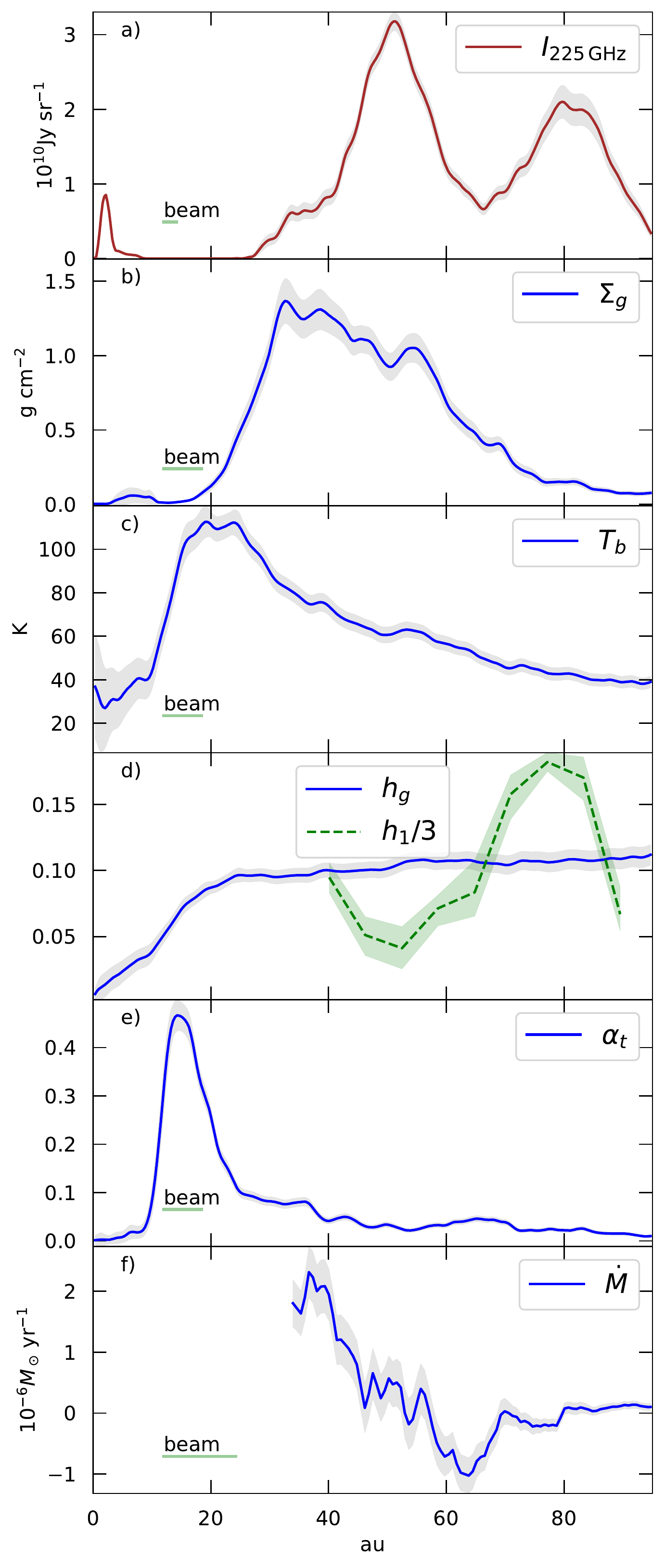}
  \caption{Radial continuum intensity profile and properties of the
    HD\,135344B  inferred from the uniform slab diagnostics with
    {\sc Slab.Line} and from the disc rotation curve: {\bf a)}
    Azimuthal average of the 225\,GHz continuum intensity profile
    extracted from the image in Fig.\,\ref{fig:HD135344B_cont}c.  {\bf
      b)} gas surface density profile $\Sigma_g(R)$, {\bf c)} gas
    temperature profile $T_b(R)$, {\bf d)} aspect ratio profile
    $h_g(R)=c_s/\Omega_K$, and comparison with the unit-optical-depth surface $h_1(R)$, {\bf
      e)} turbulence parameter $\alpha_t = (v_{\rm turb}/c_s)^2$, {\bf
      f)} accretion rate $\dot{M}(R) = -2 \pi R \Sigma_g \tilde{v}_R$.
  } \label{fig:props}
\end{figure}

The large continuum crescent in the outer ring, at $\sim$80\,au or
$\sim 0\farcs6$, is absent from the $\Sigma_g(\vec{x})$ map in
Fig.\,\ref{fig:linefit_HD135344B} (and also from the azimuthal
averages in Fig.\,\ref{fig:props}).  This may seem surprising in the
context of the dust trap interpretation, since the required local
pressure maximum should correspond to a local peak in $\Sigma_g$
\citep[][even if this local peak can be quite
  shallow]{Birnstiel2013A&A...550L...8B}. However, a similar result is
seen in MWC\,758, where the position of the dust peak does not
correspond to a maximum in $\Sigma_g$, as inferred from the CO
isotopologues \citep[][]{Boehler2018ApJ...853..162B}.  In HD\,135344B,
while it appears that the CO mass surface density is fairly
homogeneous across the disc, it may be that the CO(2-1) isotopologues
come short of reaching deep enough to sample the total mass surface
density. This could result from freezing of CO onto dust grains, or
because the continuum is optically thick, so that thermal equilibrium
between dust and gas in the denser regions would leave no net line
emission after continuum subtraction \citep[see][for a detailed
  description of this effect]{Boehler2017ApJ...840...60B}.


The radial profiles for $\Sigma_g$ and $T_b$, obtained with azimuthal
averages and a disc inclination of $i=16\,$deg, are shown in
Figs.\,\ref{fig:props}b and \ref{fig:props}c. The CO line temperatures
reach close to $\sim$115\,K, which would be the dust temperature for
water condensation, out to $\sim$25\,au. Within $\sim$20\,au,
confusion of different Keplerian velocities in the finite beam
exaggerate the turbulent velocities along an inner ring inset within
the hot inner ring seen in $T_b$, which is itself inset within the
dense ring seen in $\Sigma_g$.  The lack of signal inside the
$^{12}$CO central cavity (within 0\farcs1) yields spuriously low
values for $T_b$ because we have set a maximum value for $T_b$ of
3$\times$ the peak line brightness temperature (see
Sec.\,\ref{sec:slab.line}). Note that releasing this upper limit on
$T_b$ yields somewhat higher temperatures and somewhat lower values
for $v_{\rm turb}$ inside the central cavity, but also leads to glitches in
$T_b$ in the outer regions (albeit with little impact on the other
parameters). The best fit values for $T_b$ come close, at $\sim
-$30\%, of the radial temperature profile of the $\tau=1$ surface in
the $^{12}$CO(2-1) line as estimated from detailed thermochemical
modelling of the (3-2) CO isotopologue lines \citep[DALI model in
][]{vdMarel_2016A&A...585A..58V}. In Fig.\,\ref{fig:props}c, $T_b$
appears to be above the CO sublimation temperature of $\sim$20\,K
everywhere in the CO layer. However, colder CO in the mid-plane could
have condensed on the dust grains, where the gas phase CO abundance is
much lower
\citep[$f_{^{12}{\rm CO}} \sim 10^{-12}$,][]{vdMarel_2016A&A...585A..58V}. This is reflected in the factor
of $\sim$10 larger gas surface density derived from the CO(3-2)
isotopologues with DALI. The {\sc Slab.Line} diagnostics thus yield a
lower limit to the total gas mass, since in general $f_{^{12}{\rm CO}}
< 10^{-4}$. The global structure of $\Sigma_g(R)$ inferred from {\sc
  Slab.Line} is consistent with previous estimates based on parametric
modelling, either with the location of the sharp gas cavity edge
placed at 30\,au by \citet{vdMarel_2016A&A...585A..58V}, or with the
gradual density drop inwards used by \cite{vdM2021AJ....161...33V},
with a gap at $\sim$20\,au.

The lack of $^{12}$CO integrated intensity at the centre of the cavity
is suggestive of a very low column density of $^{12}$CO, and is
reflected in the value of $\Sigma_g$ near the star, which is
consistent with zero. This central hole could be caused by
photo-dissociation driven by UV irradiation. The $^{12}$CO integrated
intensity is so low in the disc’s innermost regions that
photo-dissociation could act down to the disc mid-plane. For this to
happen, gas surface densities $\lesssim 10^{-2}$ g cm$^{-2}$ are
necessary (see, e.g., Eq.~4 of \citealp{Baruteau2021} for
$z=0$). Interestingly, the modelling of CO ro-vibrational observations
by \citet{Carmona2014} indicates that the gas surface density in the
cavity of the HD135344B should be $\lesssim 10^{-2}$ g cm$^{-2}$ (see
the lower-right panel in their Fig.\,6), which would support the idea
that photo-dissociation could indeed be responsible for the lack of
$^{12}$CO integrated intensity inside the cavity.

%

\subsection{Rotation curve}  \label{sec:rotcurve}

The  rotation curve can be extracted from the velocity centroid
map and brings constraints on the central (stellar) mass and 
orientation \citep[][]{CasassusPerez2019ApJ...883L..41C}. From the
observations presented in Sec.\,\ref{sec:obs}, the velocity field of
the top layer in $^{12}$CO(2-1) is best traced with the
double-Gaussian moments applied to either the {\sc uvmem}-restored or
the $uv$-tapered data cubes. For conciseness we choose to
report on the $uv$-tapered version, as it allows an extension to
larger radii and is consistent with the results from the finer angular
resolutions, even at radii as small as 0\farcs25.

The 3-D rotation curve $\vec{\tilde{v}}(R)=(\tilde{v}_R(R),
\tilde{v}_\phi(R), \tilde{v}_z(R))$, in disc-centred cylindrical
coordinates where $z=0$ coincides with the mid-plane, also informs on
large scale radial and vertical flows
\citep[][]{Teague2019Natur.574..378T}. We extended the same procedure
as described in \citet[][]{CasassusPerez2019ApJ...883L..41C} to 3-D in
the {\sc ConeRot} package\footnote{publicly available at
\url{https://github.com/simoncasassus/ConeRot}}. In an axially
symmetric disc the unit opacity surface in an optically thick line
such as $^{12}$CO(2-1) can be represented by its height over the
mid-plane, $h_1(R)$. We approximate this surface by a series of cones
whose orientations are fit to the observed velocity centroid in
concentric radial domains, or `regions', which we combine by averaging
as described in \citet[][]{CasassusPerez2019ApJ...883L..41C}. The
procedure is similar to that followed by the {\sc eddy} package
\citep{Teague2019Natur.574..378T}, except that the disc orientation
along with the rotation curve are both optimised to fit $v^\circ_1$ in
each region.



A full optimisation to fit $v^\circ_1$ over the radial domain
$[0\farcs3,0\farcs7]$, with an axially symmetric model in a purely
azimuthal flow and varying the disk orientation, results in loose
constraints on the disk inclination. In initial trials we used the
{\tt tclean} datacubes, $uv$-tapered to a $0\farcs12$ beam, and
obtained $ i = 17.6^{+2.7}_{-3.2}$\,deg,
PA$=241.9^{+0.4}_{-0.5}$\,deg, and an aspect ratio
$h_1=0.28^{+0.06}_{-0.05}$ (for the unit-opacity surface), and a
systemic velocity $v_{\rm lsr}= 7.12\pm0.02\,$km\,s$^{-1}$
\citep[see][for details]{CasassusPerez2019ApJ...883L..41C}.  When
using the definitive dataset, based on the {\sc uvmem} reconstructions
(so with a 20\% narrower beam at the expense of slightly noisier
maps), we obtain $ i = 23.4^{+2.0}_{-6.2}$\,deg,
PA$=242.4^{+0.6}_{-0.7}$\,deg, $h_1=0.27^{+0.19}_{-0.08}$ and $v_{\rm
  lsr}= 7.10\pm0.02\,$km\,s$^{-1}$.

\begin{figure}
  \centering
  \includegraphics[width=\columnwidth,height=!]{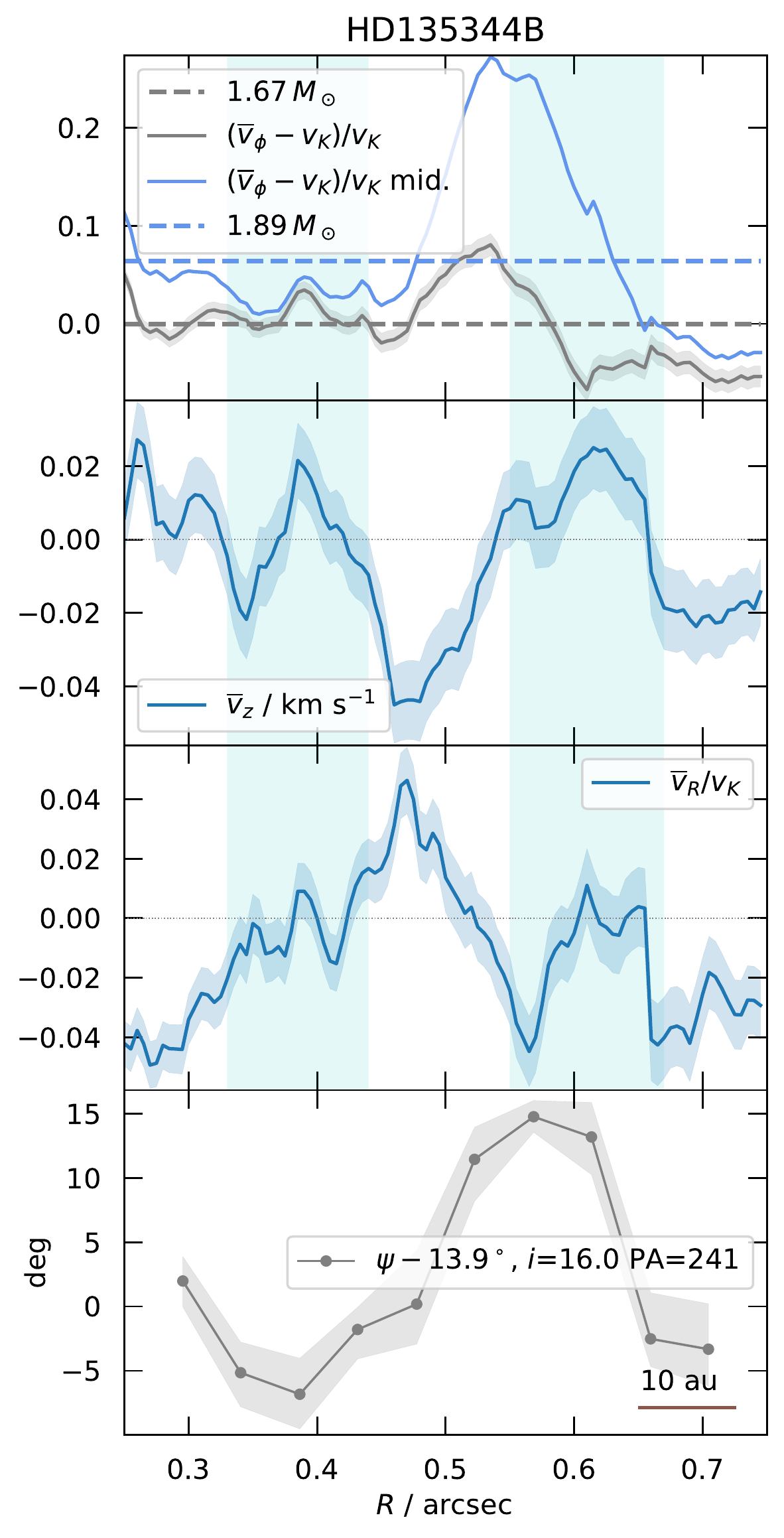}
  \caption{Rotation curve in HD\,135344B.  The regions in cyan
    correspond to the total extent of the two rings. From top to
    bottom, we show: {\bf 1)} The azimuthal rotation curve
    $\tilde{v}_\phi(R)$. The dashed horizontal lines are comparison
    Keplerian profiles with the corresponding stellar mass.  The curve
    labelled `mid' is an extrapolation to the disc mid-plane assuming
    vertical Keplerian shear.  {\bf 2)}: The vertical velocity
    component curve $\tilde{v}_z(R)$, where $\tilde{v}_z >0$ points
    away from the disk mid-plane. {\bf 3)}: The radial velocity
    component $\tilde{v}_r(R)$, where $\tilde{v}_r >0$ points away
    from the star .  {\bf 4)}: The opening angle of the cone tracing
    the unit opacity surface for $^{12}$CO(2-1).} \label{fig:rotcurve}
\end{figure}
\begin{figure*}
  \centering
  \includegraphics[width=\textwidth,height=!]{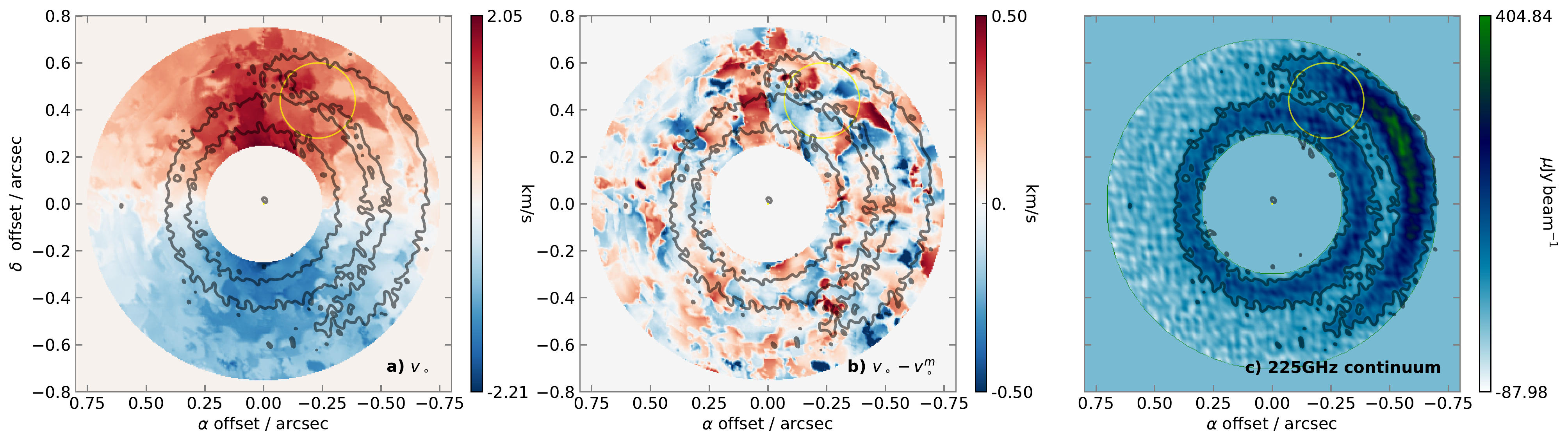}
  \caption{Face-on view of the kinematics in HD\,135344B, inferred
    using {\sc ConeRot} and the top-layer velocity centroid for the
    $^{12}$CO(2-1) $uv$-tapered datacube. {\bf a)} Deprojected
    velocity centroid (approximating the disc with a cone). The single
    contour traces the continuum image shown in c), at 20\% peak.
    {\bf b)} Difference between the observed velocity field and the
    axially symmetric model, showing essentially thermal
    residuals. Deeper data are required to ascertain the structures
    seen near the position of the filament. {\bf c)} Comparison with a
    face-on view of the continuum, deprojected as a thin disc with
    inclination $i=16$\,deg. In all images the yellow circle
    highlights the position of the putative filament.  } \label{fig:faceonkine}
\end{figure*}

We therefore opted to fix the inclination to $i=16\,$deg, which yields
dynamical stellar masses that are consistent with the photospheric
data \citep[][see below in this
  Section]{Wichittanakom2020MNRAS.493..234W}, and set PA$=241.9$. We
then proceeded to optimise the aspect ratio and the rotation curve in
11 radial bins over $[0\farcs25,0\farcs75]$, which produced the 3-D
rotation curve shown in Fig.\,\ref{fig:rotcurve}. The sign convention
we follow is such that $\tilde{v}_z >0$ and $\tilde{v}_R >0$ points
away from the disk mid-plane and from the star, as in an outflow. A
face-on view of the disc using this geometry is shown in
Fig.\,\ref{fig:faceonkine}. The deviations from the axially symmetric
flow appear to be thermal, although more pronounced in the region
around the possible filament seen in the continuum. Deeper data are
required to discuss these velocity deviations.


The dynamical stellar mass, fit to the tangential component
$\tilde{v}_\phi(R)$, is $1.67 \pm 0.04 < M_\star /M_\odot < 1.89 \pm
0.04$. The lower limit stems from assuming perfect cylindrical
rotation, while the upper limit corresponds to Keplerian vertical
shear.  This stellar mass is consistent with that measured from the
photospheric spectrum \citep[][]{Wichittanakom2020MNRAS.493..234W},
$M_\star = 1.67\,^{+0.18}_{-0.16} M_\odot$.

In a rotation curve fit the resulting stellar mass is quite sensitive on inclination $i$:
\begin{equation}
M_\star \propto  R\times \tilde{v}^2_\phi(R) \propto  \frac{1}{\cos(i)} \times \frac{1}{\sin^2(i) }, \label{eq:Mstar}
\end{equation}
if the extracted rotation curve and disk aspect ratio are kept equal
(aside from the $\sin(i)$ factor). For comparison, if we fix
$i=24\,$deg (as inferred from the continuum), an application of {\sc
  ConeRot} gives $0.91 \pm 0.04 < M_\star /M_\odot < 0.98 \pm 0.04$
(the lower limit would be $0.92\,M_\odot$ in an extrapolation of the
stellar mass value from $i=17.6$\,deg using Eq.\,\ref{eq:Mstar}).  In
turn, if we fix inclination to 9.8\,deg, the measurement with the
smallest error bars in \citet[][their Band\,7
  case]{Cazzoletti2018A&A...619A.161C}, then $4.9 \pm 0.04 < M_\star
/M_\odot < 6.8 \pm 0.04$ (the lower limit from Eq.\,\ref{eq:Mstar}
would be $4.9\,M_\odot$).

If the tentative sub-mm continuum filament traces a planetary wake,
one may expect the planet in the sub-mm annular gap to also carve a
dip in the gas surface density, whose signature may be observable in
$^{12}$CO if the planet is massive enough. The opening angle of the
cone tracing the $^{12}$CO(2-1) unit opacity surface corresponds to an
aspect ratio of $h_1 \sim 0.1$ at $\sim 0\farcs4$ separation (i.e. in
the sub-mm annular gap), compared to $h_1>0.2$ beyond the annular gap
(see Fig.~\ref{fig:rotcurve}).

\section{Discussion} \label{sec:discussion}



As noted in Sec.\,\ref{sec:obs}, the same 2-armed grand-design spiral
that characterises the near-IR scattered-light images also stands out
in the $^{12}$CO(2-1) velocity-integrated intensity and in the line
velocity dispersion (or second-order moment). However, the spiral
pattern is absent in the peak intensity map
(Fig.\,\ref{fig:sgaussmoms}). The absence of a counterpart in peak
intensity suggests that the spiral structure is not due to a local
enhancement in temperature, but is instead due either to a strong
modulation in gas surface density, or to enhanced velocity dispersion.
The lack of a conspicuous spiral pattern in the rarer isotopologue
maps, which are optically thinner, suggests that the spirals probably
do not correspond to enhanced surface densities. These arguments
tentatively support the case of ``turbulent spirals'', i.e. that
turbulence, or superposed velocity components along the line-of-sight,
is strong in this disc and particularly so in the spirals.

The uniform-slab diagnostics obtained with {\sc Slab.Line} show that
the grand-design 2-armed spiral in velocity-integrated intensity is neither
seen in the gas surface density $\Sigma_g(\vec{x})$ nor in the gas
temperature $T_b(\vec{x})$. Instead, it is reflected in the turbulent
velocity map $v_{\rm turb}(\vec{x})$. This is surprising as most
attempts to constrain the level of turbulence in protoplanetary discs
have resulted in upper limits, except in DM\,Tau \citep[see][and
  references therein]{Flaherty2020ApJ...895..109F}, where values for
$v_{\rm turb}$ are similar as reported here for HD\,135344B.


In the $\alpha$-viscosity model \citep[][]{1973A&A....24..337S}
viscosity is related to the thermal disc structure by a dimensionless
parameter, $\alpha_t$. We can re-formulate our results by calculating
this parameter locally, by comparing the turbulent velocities with the
local sound speed, $\alpha_t \approx (v_{\rm turb}/c_s)^2$.  The
uniform-slab diagnostic yields the gas temperature $T_b(\vec{x})$,
from which we obtain the sound speed $c_s =\sqrt{ \gamma k_B T_b/ (\mu
  m_p)}$, with an adiabatic index $\gamma=1.4$ and a molecular weight
$\mu=2.3$. Fig.\,\ref{fig:props} shows that $\alpha_t(R)$ reaches
values\footnote{the steep rise towards the origin corresponds to the
spurious inner ring in $v_{\rm turb}$ that is due to weak signal and confusion, as
discussed in Sec.\,\ref{sec:lines}} of $\sim 0.05$.

The inferred values for $\alpha_t$ seem very high, especially in the
vortex interpretation of the large crescent modulating the outer ring.
This is because very low levels of `alpha' viscosity , of order
$\alpha \sim 10^{-4}$--$10^{-3}$, are necessary to develop the
Rossby-wave instability and explain the outer ring crescent with an
anticyclonic vortex
\citep[e.g.][]{Barge_Sommeria_1995A&A...295L...1B,LyraLin2013ApJ...775...17L,
  Zhu_Stone_2014ApJ...795...53Z, ZB2016MNRAS.458.3918Z}. The turbulent
velocity map in HD\,135344B is thus unlikely to actually trace the
source of viscosity in the disc, but may instead reflect the
superposition of several velocity components, or superposed laminar flows along
the line of sight.

Interestingly, the rotation curve of HD\,135344B points at strong
accretion towards the star, especially inside the inner dust ring (so
$< 0\farcs4$). The radial velocity component, with a peak in absolute
value $v_R =-0.34\pm0.05$\,km\,s$^{-1}$ (Fig.\,\ref{fig:rotcurve}),
comes close to the sound speed, which ranges from
$\sim0.64\pm0.05$\,km\,s$^{-1}$ to
$c_s \sim 0.44\pm0.04 $\,km\,s$^{-1}$ over the radial domain plotted
in Fig.\,\ref{fig:props}. Transonic velocities are theoretically
expected in the cavity of transition discs where magnetised winds
result from thermal (photoevaporative) winds threaded by magnetic
field lines, which exert a torque on the gas remaining in the wind and
drive accretion \citep[][]{Wang2017ApJ...835...59W}. The likelihood of
photoevaporative winds and net poloidal fields in discs is
theoretically established \citep[e.g.][ and references
therein]{Ercolano2017RSOS....470114E}.

We can estimate the corresponding accretion rate with $\dot{M}(R) = -2
\pi R \Sigma_g \tilde{v}_R$. As illustrated in Fig.\,\ref{fig:props}f,
the peak accretion rate is $\dot{M} =
(2\pm0.1)\times10^{-6}\,M_\odot\,$yr$^{-1}$. This rate is 40 times
larger than the stellar accretion rate of
$5\times10^{-8}\,M_\odot\,^{-1}$
\citep[]{Fairlamb2015MNRAS.453..976F}, and would deplete the disc mass
sampled by CO(2-1), or $\sim 1.6\times 10^{-3}\,M_{\odot}$, in $\sim
800$\,yr (the fraction of the material accreted by a potential planet
inside the ring is small). It may be that the mass reservoir is not
sampled by CO, but even a very massive disc, with $0.1\,M_{\odot}$,
would still be much too short lived. The system may, perhaps, be
observed in a transient phase, or an important fraction of the
material being accreted inside $\sim$50\,au will eventually be
expelled in a wind.

A similar issue as the disc depletion timescale also arises with the
diverging accretion rates near 50\,au, so right on the inner dust ring
(see Fig.\,\ref{fig:props}f). Since the accretion rate is negative
right outside the ring, and reaches $\sim (-1\pm0.2)\times 10^{-6}
M_\odot\,$yr$^{-1}$ at $\sim 60\,$au, a gap would quickly develop at
the location of the ring.

Another solution to both the very high values for $\alpha_t$, and the
very high apparent accretion rate, is that accretion is restricted
only to the surface layer, where the $^{12}$CO(2-1) line
originates. This would be a similar situation as that of meridional
flows at the edges of a protoplanetary gap, but here in the case of
the outer edge of the central cavity. The mid-plane, enclosing most of
the disc mass, may instead be almost in pure azimuthal rotation, with
at least a factor of 10 smaller radial velocity than in the surface so
as to reconcile the disc accretion rate with the stellar accretion. If
this is the case, then the origin of the line broadening would indeed be the
superposition of laminar flows along the line of sight, rather than
genuine turbulence. The possibility of a strongly decreasing gradient
for the radial velocity component towards the mid-plane can be tested
with an extension of the rotation curve analysis from
Sec.\,\ref{sec:rotcurve} to deeper CO isotopologue data. Theoretical
models indeed suggest that both in the case of MRI-driven or
wind-driven accretion, the active (accreting) layer in discs is
expected to be confined to the surface
\citep[e.g. ][]{Mohanty2013ApJ...764...65M, Bai2016ApJ...821...80B}.

\section{Conclusions}

The HD\,135344B disc is especially interesting in the context of
planet-disc interactions. Here we reported on new ALMA observations,
with unprecedented angular resolutions in this source. The data
correspond to a partial delivery with about $1/10$ of the total
integration, but nonetheless reveal interesting aspects of this
disc, whose statistical significance will be further assessed pending delivery of the full dataset:


\begin{itemize}
\item A thin filament in the continuum image at 225\,GHz crosses the
  gap between the outer and inner rings (Fig.\,\ref{fig:HD135344B_cont}). 
  Although its median intensity is at $4$ times the noise level, confirmation 
  of this filament requires a second epoch and deeper observations. The filament is
  found at the same location as the putative protoplanet proposed to
  be driving one of the spirals by \citet{Muto2012ApJ...748L..22M},
 and almost co-aligned with a local twist in the IR spiral arm (same PA but 
 radially shifted; Fig.\,\ref{fig:spiral_fit}c). We suggest that the radial shift is due to
 the curling of the spiral towards the star at the disc surface \citep{Zhu2015}.
  
\item The $^{12}$CO(2-1) velocity-integrated and dispersion maps trace
  the same spiral seen in scattered light, and characteristic of
  HD\,135344B (see Figs.\,\ref{fig:12COuvmem}, \ref{fig:sgaussmoms}
  and \ref{fig:spiral_fit}c). The CO spiral is modulating an extended
  disc, and its arm-inter-am contrast is much shallower than in the
  near-IR. It is not affected by shadowing from a possible tilted
  inner disc. The spiral pattern and extended disc are absent from the
  peak intensity map, suggesting that the origin of the line
  broadening is turbulence rather than a temperature wave.

\item The physical conditions and line-of-sight turbulent broadening
  $v_{\rm turb}$ inferred from the uniform-slab and LTE approximations
  confirm that the extended disc is axially symmetric in the surface
  density of the CO layer, while $v_{\rm turb}$ follows the IR spiral (Fig.\,\ref{fig:linefit_HD135344B}).

\item The magnitude of $v_{\rm turb}$ in this disc is very large, and
  close to $\sim$22\% sonic. The corresponding viscosity is $\alpha_t
  \sim 0.05$ (Fig.\,\ref{fig:props}e), which is much higher
  than standard values and suggests that the line broadening stems
  from superposed laminar flows rather than genuine turbulence.

\item The disc rotation curve points at an inclination of
  $\sim$16\,deg, which is consistent with estimates of the central
  star mass from the photospheric spectrum. If this inclination is
  correct, the inner ring in continuum emission is quite eccentric,
  with $e = 0.14\pm0.04$, as the inclination required to circularise
  it is $i=23.8^{+2.5}_{-2.6}$\,deg
  (Fig.\,\ref{fig:faceon_eccentric}).

\item If the gaseous disc is axially symmetric, then its 3-D rotation
  curve (Fig.\,\ref{fig:rotcurve}), including radial and vertical
  axially symmetric flows, points at strong accretion inside the inner
  dust ring, within $\sim 0\farcs4$, with a radial velocity of up to
  $v_R = -0.34\pm0.05$\,km\,s$^{-1}$. The corresponding mass accretion
  rate is $ \dot{M} = (2\pm0.2)\times10^{-6}\,M_\odot\,$yr$^{-1}$
  (Fig.\,\ref{fig:props}f), which may be reconciled with the $\sim$40
  times lower stellar accretion rate if only the surface layers are
  undergoing accretion.

\end{itemize}

\section*{Acknowledgements}

We thank Dr. Adele Plunkett and the NAASC ALMA staff for the reduction
and partial delivery of the ALMA data presented here. We also thank
Takayuki Muto, the referee, for a thorough review and constructive
comments that improved this article. This paper makes use of the
following ALMA data: ADS/JAO.ALMA\#{\tt 2018.1.01066.S}. ALMA is a
partnership of ESO (representing its member states), NSF (USA) and
NINS (Japan), together with NRC (Canada), MOST and ASIAA (Taiwan), and
KASI (Republic of Korea), in cooperation with the Republic of
Chile. The Joint ALMA Observatory is operated by ESO, AUI/NRAO and
NAOJ. S.C., S.P.  and L.A.C., acknowledge support from Agencia
Nacional de Investigaci\'on y Desarrollo de Chile (ANID) given by
FONDECYT Regular grants 1211496, 1191934 and 1211656. V.C., C.P.  and
D.J.P. acknowledge funding from the Australian Research Council via
FT170100040 and DP180104235. M.C. acknowledges support from ANID
PFCHA/DOCTORADO BECAS CHILE/2018-72190574.  N.M. acknowledges support
from the Banting Postdoctoral Fellowships program, administered by the
Government of Canada. B.E. acknowledges the support by the DFG Cluster
of Excellence "Origin and Structure of the Universe” and of the DFG
Research Unit ``Transition Disks'’ grants FOR 2634/1, ER 685/8-1, and
ER 685/9-1. A.J. acknowledges support from FONDECYT project 1210718,
and ANID - Millennium Science Initiative - ICN12\_009.  V.F.'s
postdoctoral fellowship is supported by the Exoplanet Science
Initiative at the Jet Propulsion Laboratory, California Institute of
Technology, under a contract with the National Aeronautics and Space
Administration (80NM0018D0004). M.R. acknowledges support from the FWO
research program under project 1280121N.

\section*{Data Availability}

The full ALMA dataset for project {\tt 2018.1.01066.S} will be available on
the ALMA archive, at the term of the proprietary period. Meanwhile, the reduced ALMA data underlying this article
is available upon reasonable request to the corresponding author.  The
original analysis packages that sustain this work are publicly
available at the following URLs:\\
\url{https://github.com/simoncasassus/GMoments}\\
\url{https://github.com/simoncasassus/MPolarMaps}\\
\url{https://github.com/simoncasassus/ConeRot}\\
\url{https://github.com/simoncasassus/Slab}


\section*{Author Contributions}

S.C: imaging, analysis, write-up, telescope proposal, software
development: {\sc GMoments}, {\sc MPolarMaps}, {\sc ConeRot} and {\sc
  Slab.Line}. V.C: telescope proposal, Sec.\,\ref{sec:spirals} on
spiral fits.  M.C.: {\sc uvmem} imaging and software development.
S.P.: IRDAP reduction, telescope proposal.  P.W., C.B., D.P:
planet-disc hydrodynamical context.  B.E.: disc-wind context. N.vdM.:
literature on HD\,135344B.  R.D.: Estimate of ring eccentricity. A.J:
advice on {\sc Slab.Line}. All authors commented on the manuscript.




\input{rep_C6_HD135344B.bbl}




\appendix

\section{Channel maps} \label{sec:channelmaps}

\begin{figure*}
  \centering
  \includegraphics[width=0.9\textwidth,height=!]{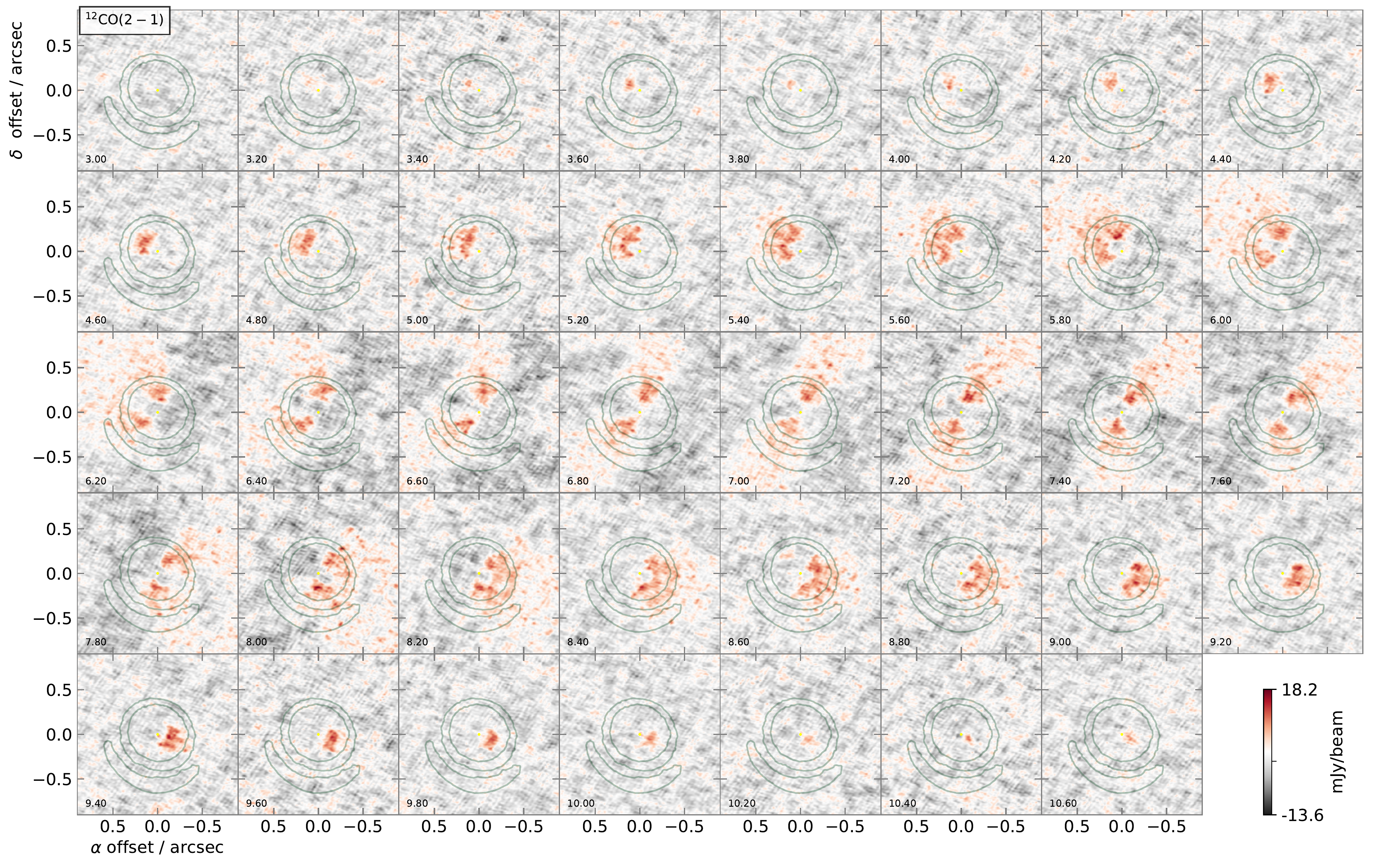}
  \caption{Channel maps from the {\tt tclean}  $^{12}$CO(2-1) datacube. The continuum from Fig.\,\ref{fig:HD135344B_cont}a is outlined in contours. The beam is $0\farcs054\times  0\farcs040 / 89.4$\,deg.   } \label{fig:HD135344B_12CO21_channels}
\end{figure*}

\begin{figure*}
  \centering
  \includegraphics[width=0.9\textwidth,height=!]{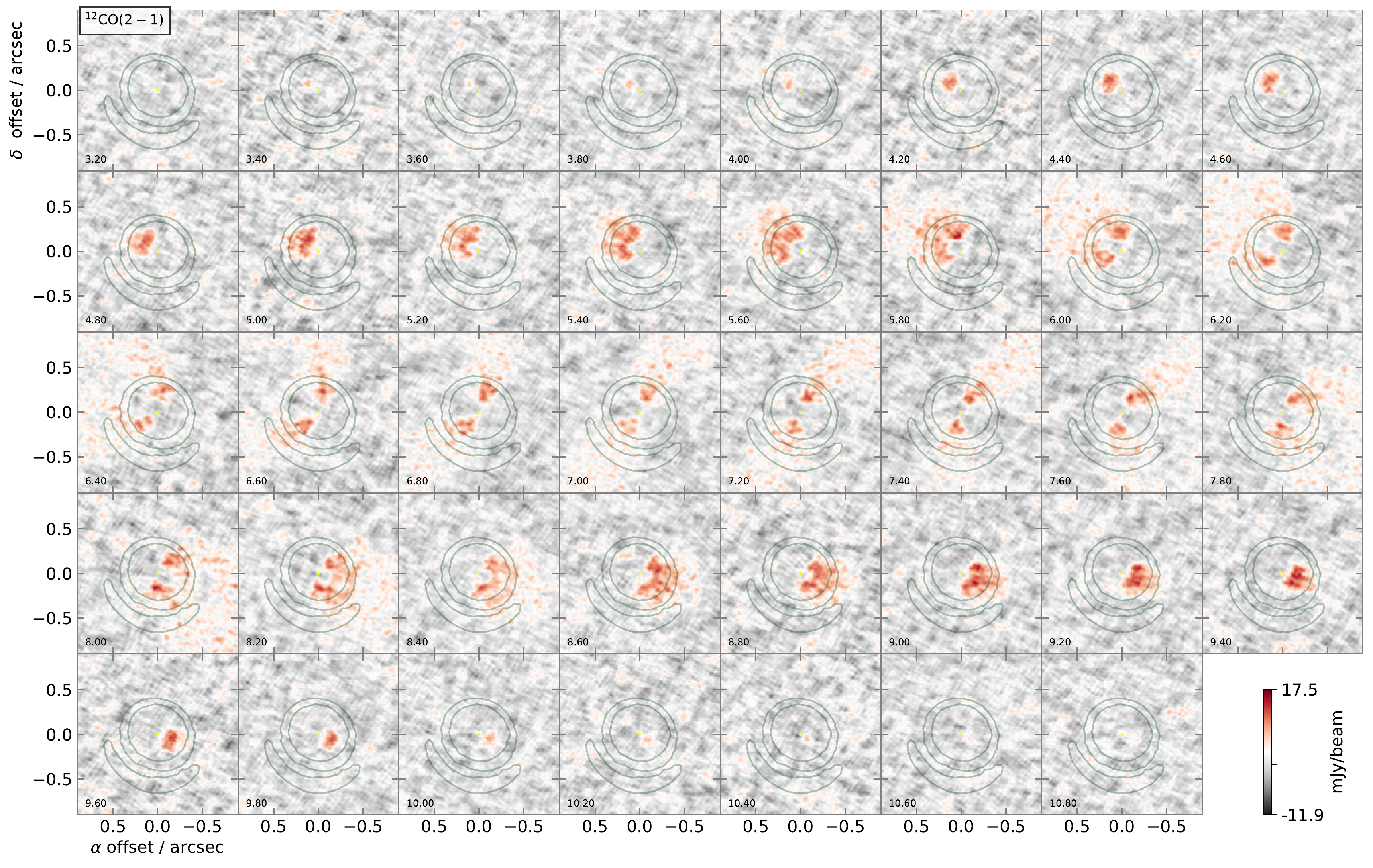}
  \caption{Channel maps from the {\sc uvmem}  $^{12}$CO(2-1) datacube. The continuum from Fig.\,\ref{fig:HD135344B_cont}a is outlined in contours. The beam is $0\farcs054\times  0\farcs040 / 89.4$\,deg.  } \label{fig:HD135344B_12CO21_channels_uvmem}
\end{figure*}

\begin{figure*}
  \centering
\includegraphics[width=\textwidth,height=!]{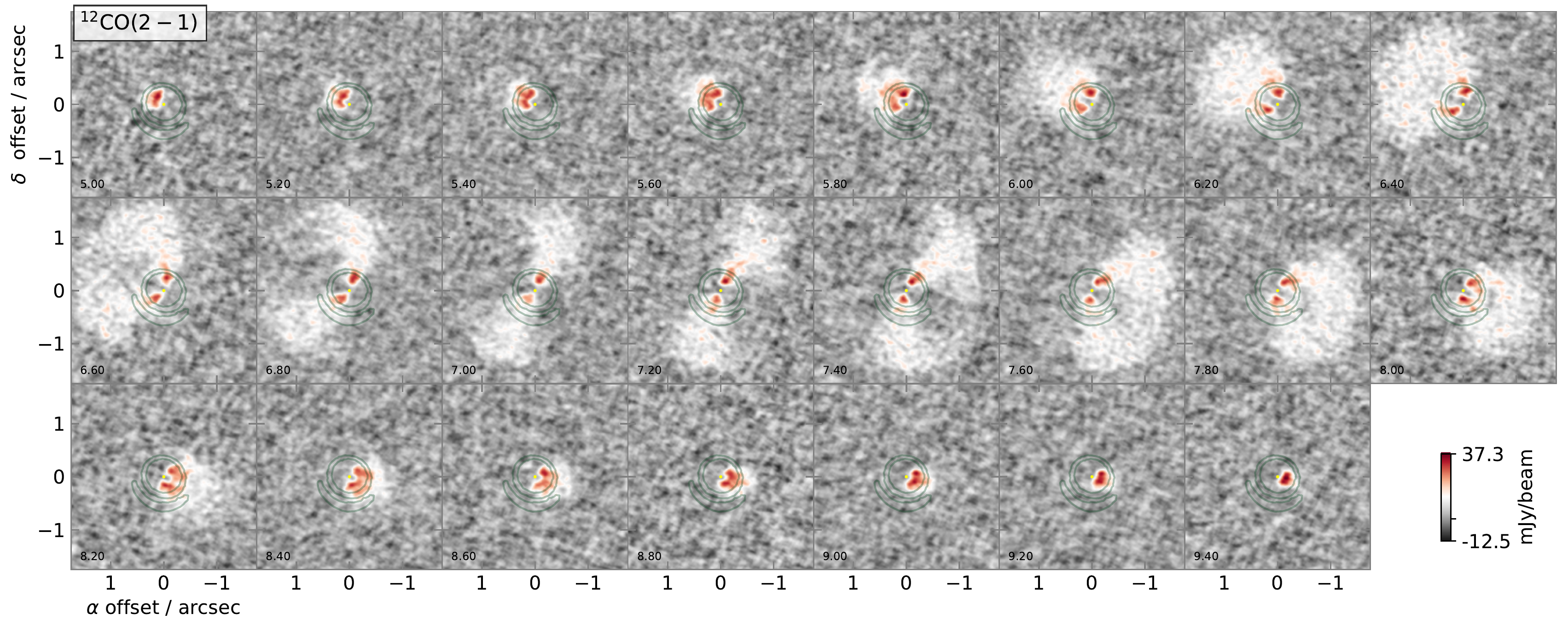}
\includegraphics[width=\textwidth,height=!]{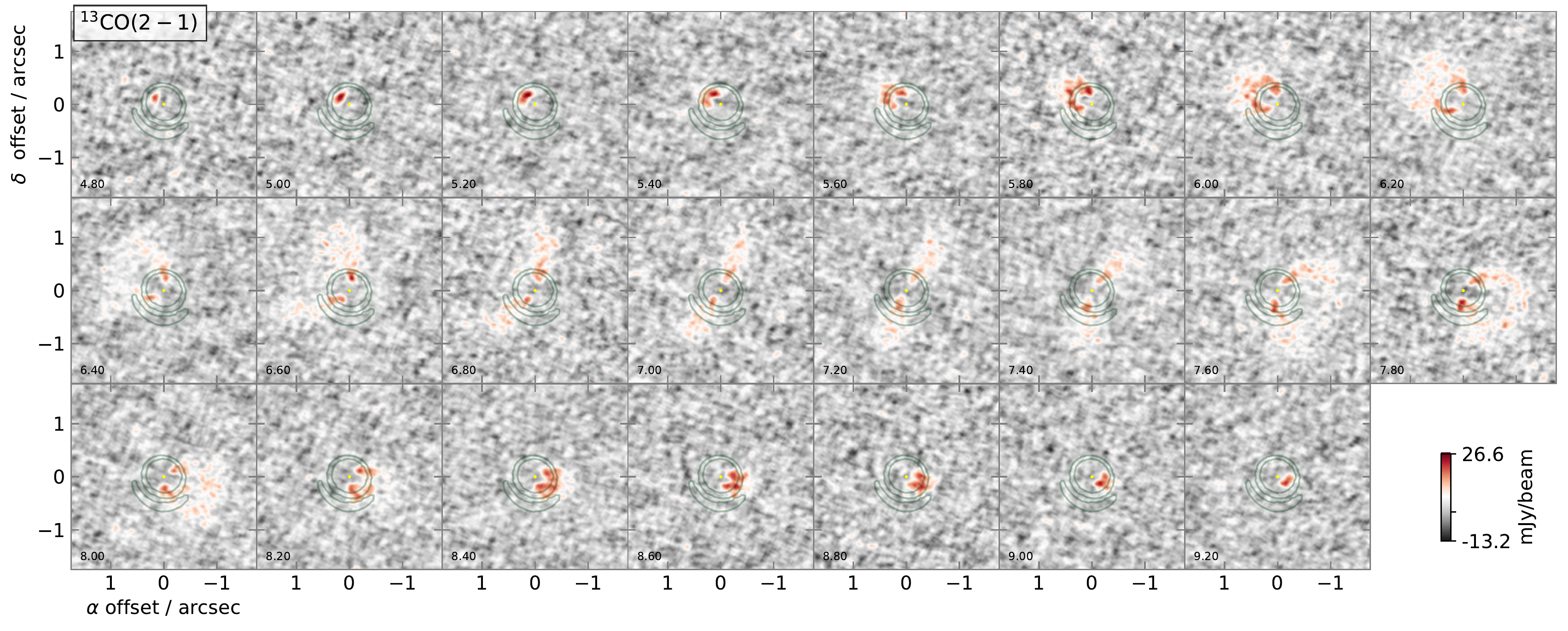}
\includegraphics[width=\textwidth,height=!]{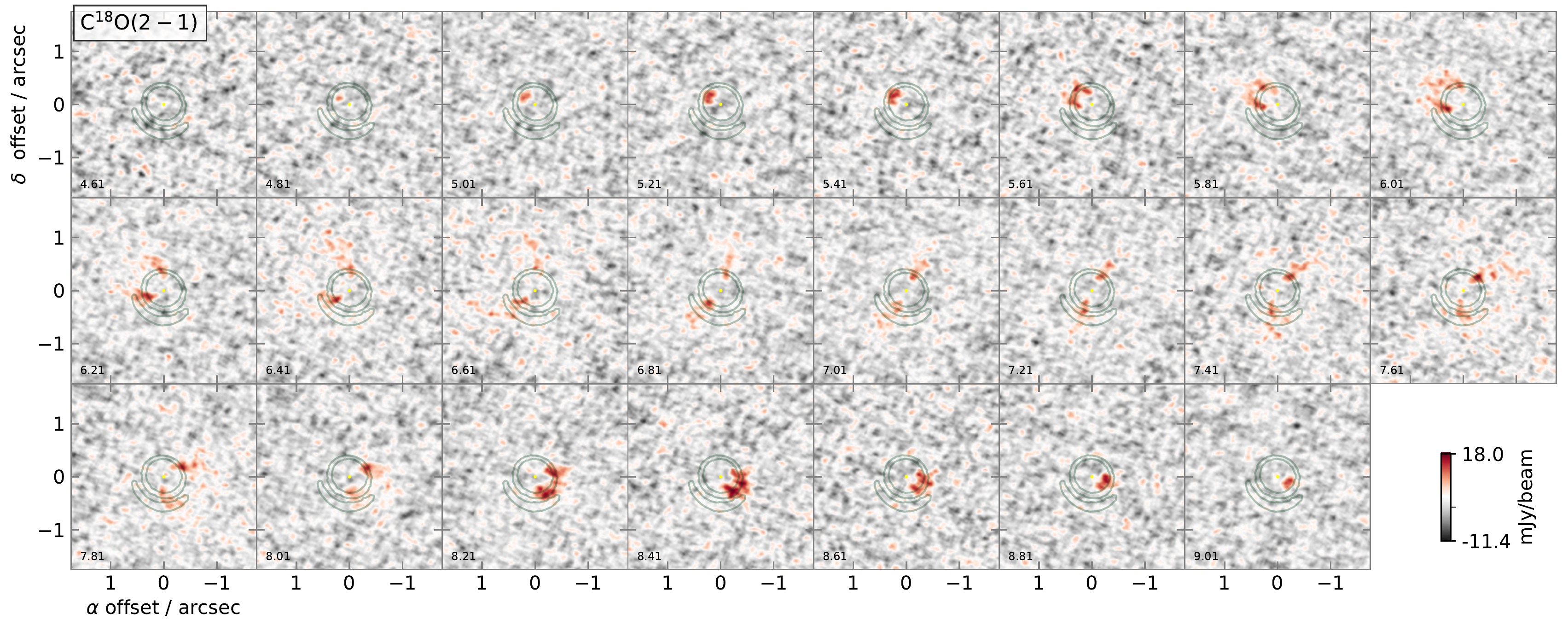}
\caption{Channel maps from the {\sc uvmem}  and $uv-$tapered datacubes for $^{12}$CO(2-1), $^{13}$CO(2-1) and C$^{18}$O(2-1). The continuum from Fig.\,\ref{fig:HD135344B_cont}a is outlined in contours. The beam is
$0\farcs100\times  0\farcs081 / 0$\,deg.  } \label{fig:HD135344B_12CO21_channels_uvmem_uvtaper}
\end{figure*}

Channels maps for the $^{12}$CO(2-1) data are shown in
Fig.\,\ref{fig:HD135344B_12CO21_channels} for the {\tt tclean}
imaging, and in Fig.\,\ref{fig:HD135344B_12CO21_channels_uvmem} for
the {\sc uvmem} image restoration. Both reconstructions used Briggs
weighting to restore the  visibility data, with a robustness parameter of
2. The {\sc uvmem} channels maps reach the same thermal noise as {\tt
  tclean}, but with less extended negatives in the channels which
correspond to emission which covers larger angular scales (near the
systematic velocity). The channel maps for the $uv$-tapered versions
of the {\sc uvmem} restorations, and for all isotopologues, are shown
in Fig.\,\ref{fig:HD135344B_12CO21_channels_uvmem_uvtaper}.


%
%
%
%
%
%

\section{disc orientation from the continuum}

Disc orientation is often inferred from continuum data under the
assumption of axial symmetry, either directly from visibility data
\citep[e.g.][]{Jennings2020MNRAS.495.3209J}, or simply by fitting
projected elliptical Gaussians in the image plane. Here we follow an
image plane approach, under the assumption of axial symmetry 
for a thin disc (with null aspect ratio).  We minimise the variance in the
radial profile for the continuum intensity $I_\nu(R,\phi)$,
\begin{equation}
  \chi^2_{\rm var} = \frac{1}{I_{\rm noise}^2}  \sum_{l=l_1}^{l=l_2} \sigma^2_\phi(R_l;{\rm PA}, i,\Delta \alpha,\Delta \delta), \label{eq:contchi2}
\end{equation}
where $I_{\rm noise}$ is the thermal noise in the image, and where
$\sigma^2_\phi$ is the azimuthal variance of intensities,
\begin{equation}
  \sigma^2_\phi(R_l)  = \frac{1}{N _\phi}\sum_{k=0}^{N_\phi -1} \left( I_\nu(R_l,\phi_k)   - \langle  I_\nu(R_l) \rangle \right)^2, 
\end{equation}
with
\begin{equation}
\langle  I_\nu(R_l) \rangle = \frac{1}{N _\phi}\sum_{k=0}^{N_\phi -1} I_\nu(R_l,\phi_k).
\end{equation}
The radial profile $\langle I_\nu(R) \rangle$ and the variance
$\sigma^2_\phi(R)$ profile  depend on the disc position angle,
inclination, and choice of origin for the (standard) polar expansion.

\begin{figure*}
\begin{center}
  \includegraphics[width=0.7\textwidth,height=!]{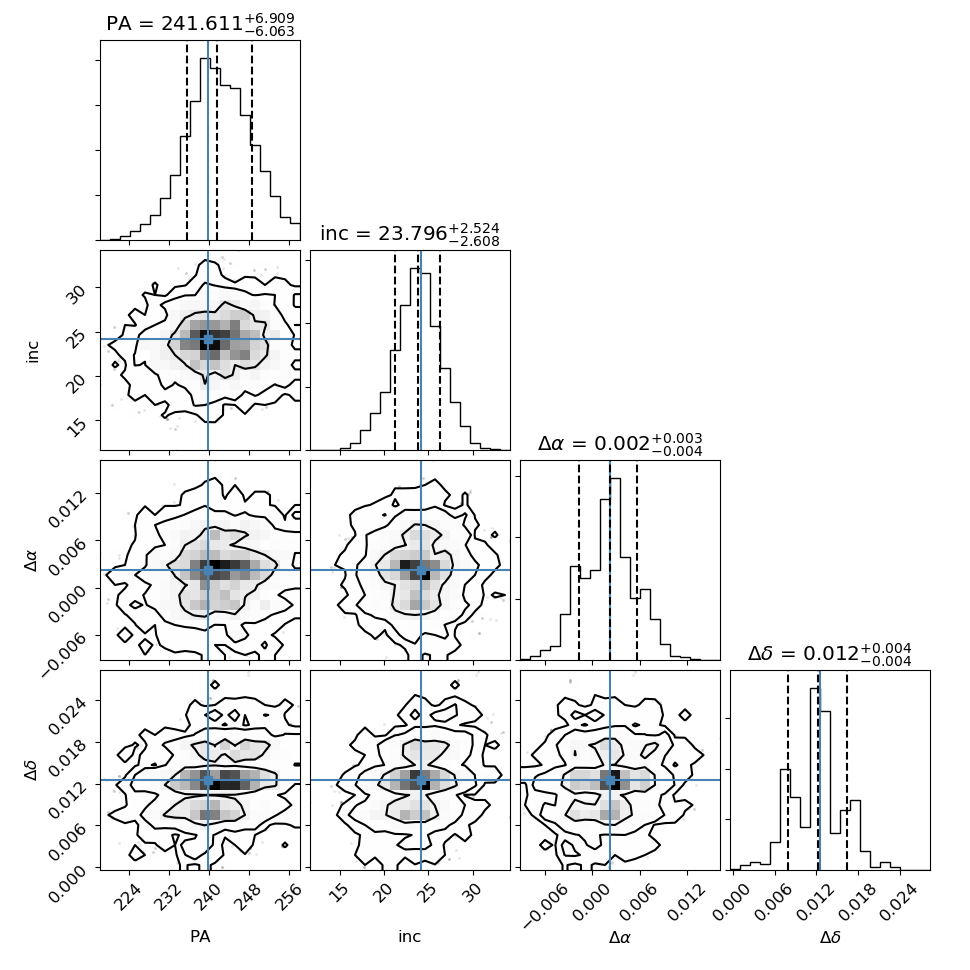}
\end{center}
\caption{Corner plot from the $\chi^2_{\rm var}$ optimisation of the
  orientation parameters in HD\,135344B using the {\tt tclean}
  continuum image at 225\,GHz continuum from
  Fig.\,\ref{fig:HD135344B_cont}b \label{fig:corner_cont}}
\end{figure*}

The minimisation of $\chi^2_{\rm var}({\rm PA}, i,\Delta \alpha,\Delta
\delta)$ in Eq.\,\ref{eq:contchi2} is carried out with the {\sc emcee}
package \citep[][]{emcee2013PASP..125..306F}, with flat priors, and
with 600 iterations and 40 walkers. The resulting posterior
distributions are summarised in Fig.\,\ref{fig:corner_cont} for an
application to HD\,135344B. This strategy to infer the disc
orientation is implemented using Python in the {\sc MPolarMaps} package, and
is publicly available at
\url{git@github.com:simoncasassus/MPolarMaps.git}.

\section{Statistical analysis of the line diagnostics} \label{sec:slab.line}

\begin{figure*}
  \centering
  \includegraphics[width=0.8\textwidth,height=!]{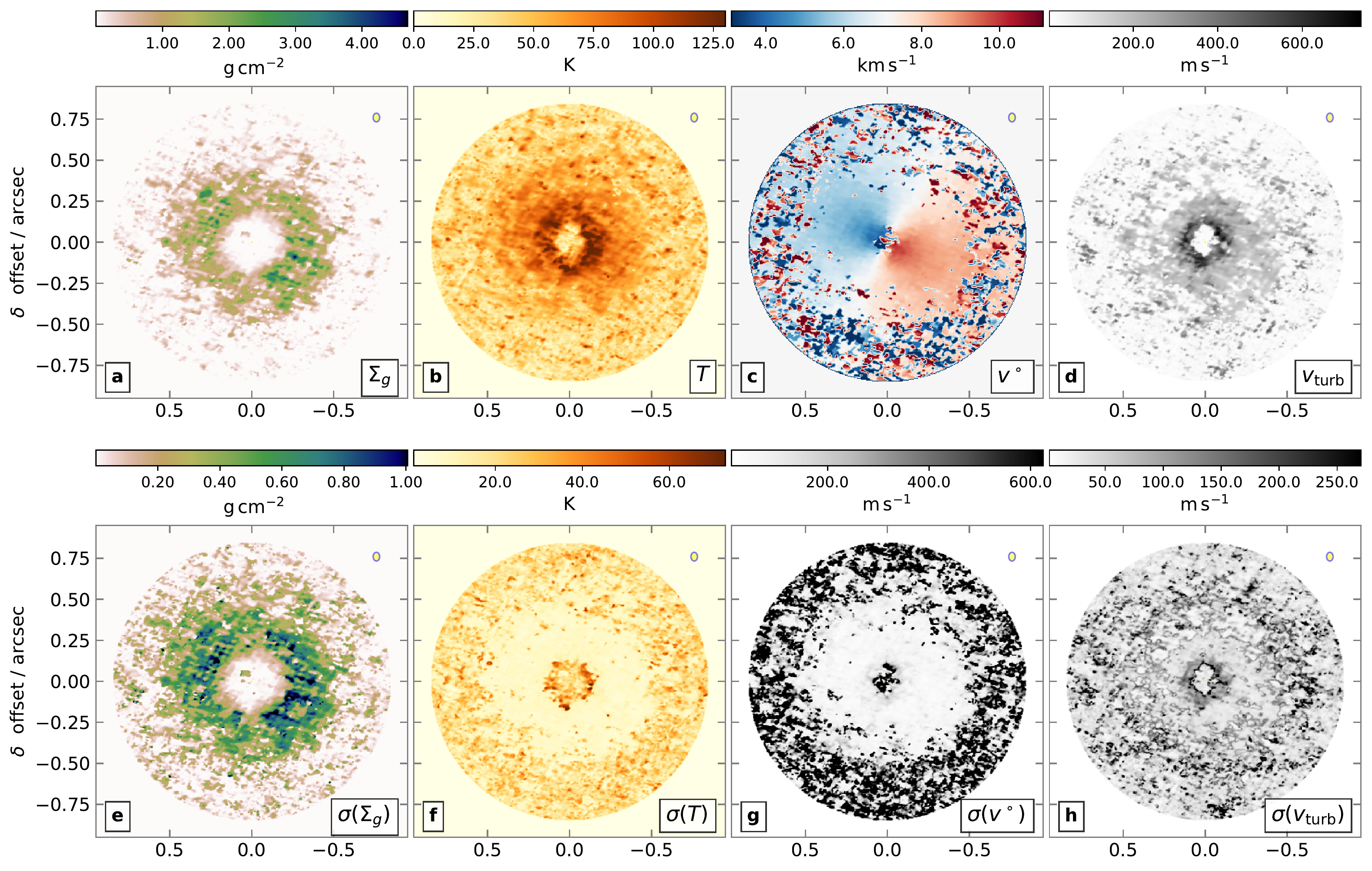}
  \caption{Expectation values and associated 1$\sigma$ errors for an application of {\sc Slab.Line} to the original CO(2-1) isotopologues data (no taper). } \label{fig:slabline}
\end{figure*}

\begin{figure*}
  \centering
  \includegraphics[width=0.8\textwidth,height=!]{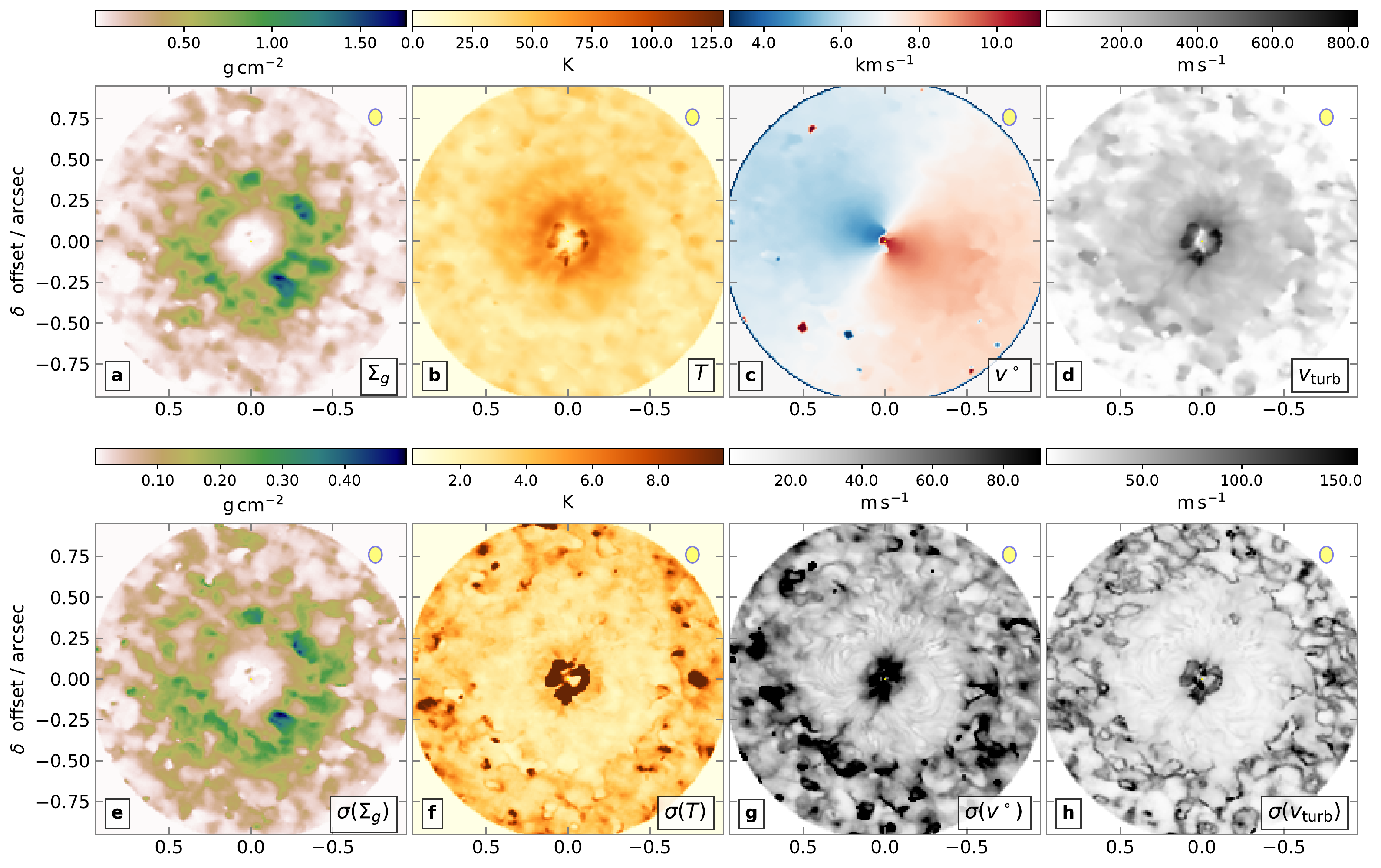}
  \caption{Same as Fig.\,\ref{fig:slabline} but for the $uv$-tapered data  } \label{fig:slabline_uvtaper}
\end{figure*}

\begin{figure*}
  \centering
  \includegraphics[width=\columnwidth,height=!]{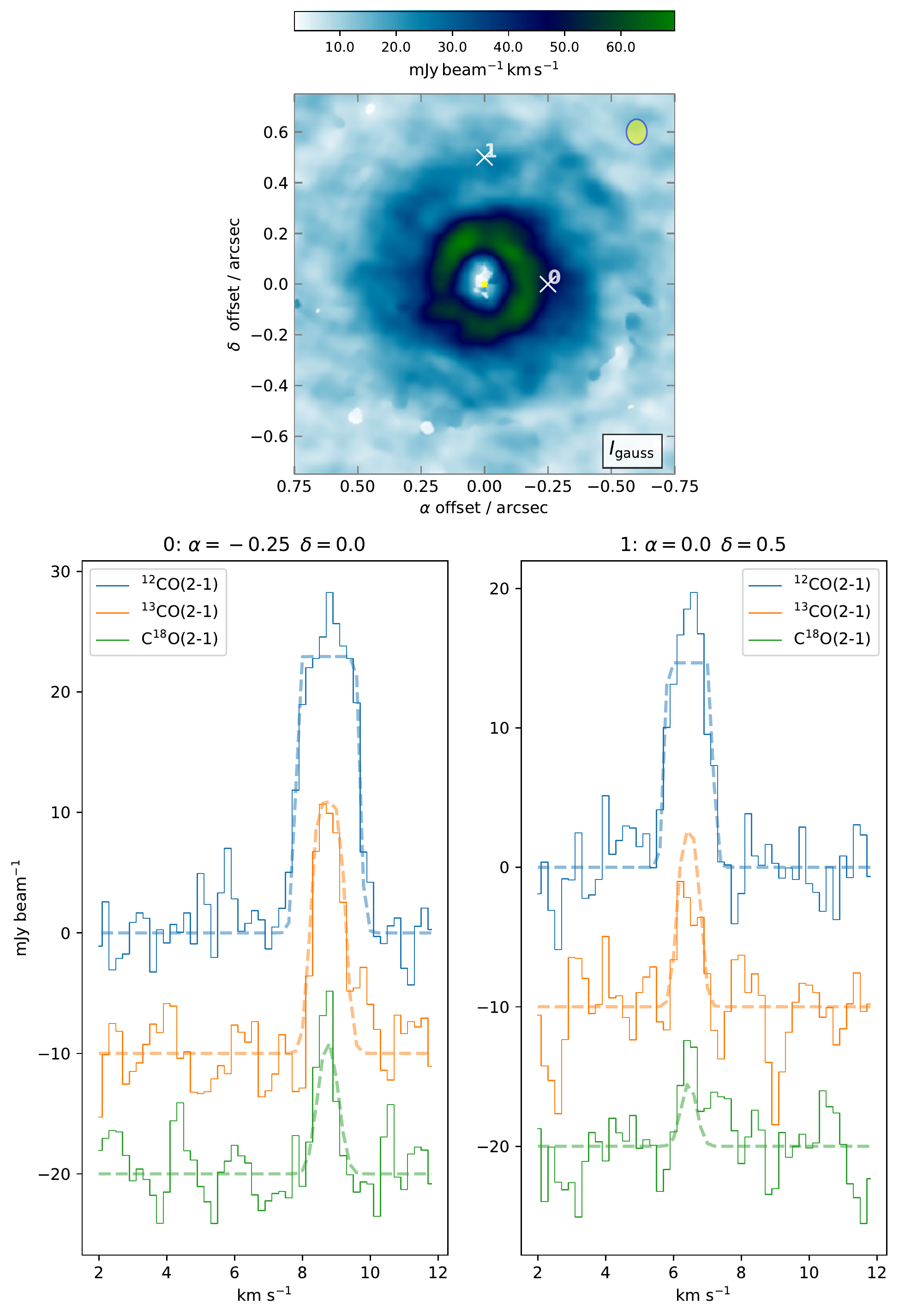}
  \includegraphics[width=\columnwidth,height=!]{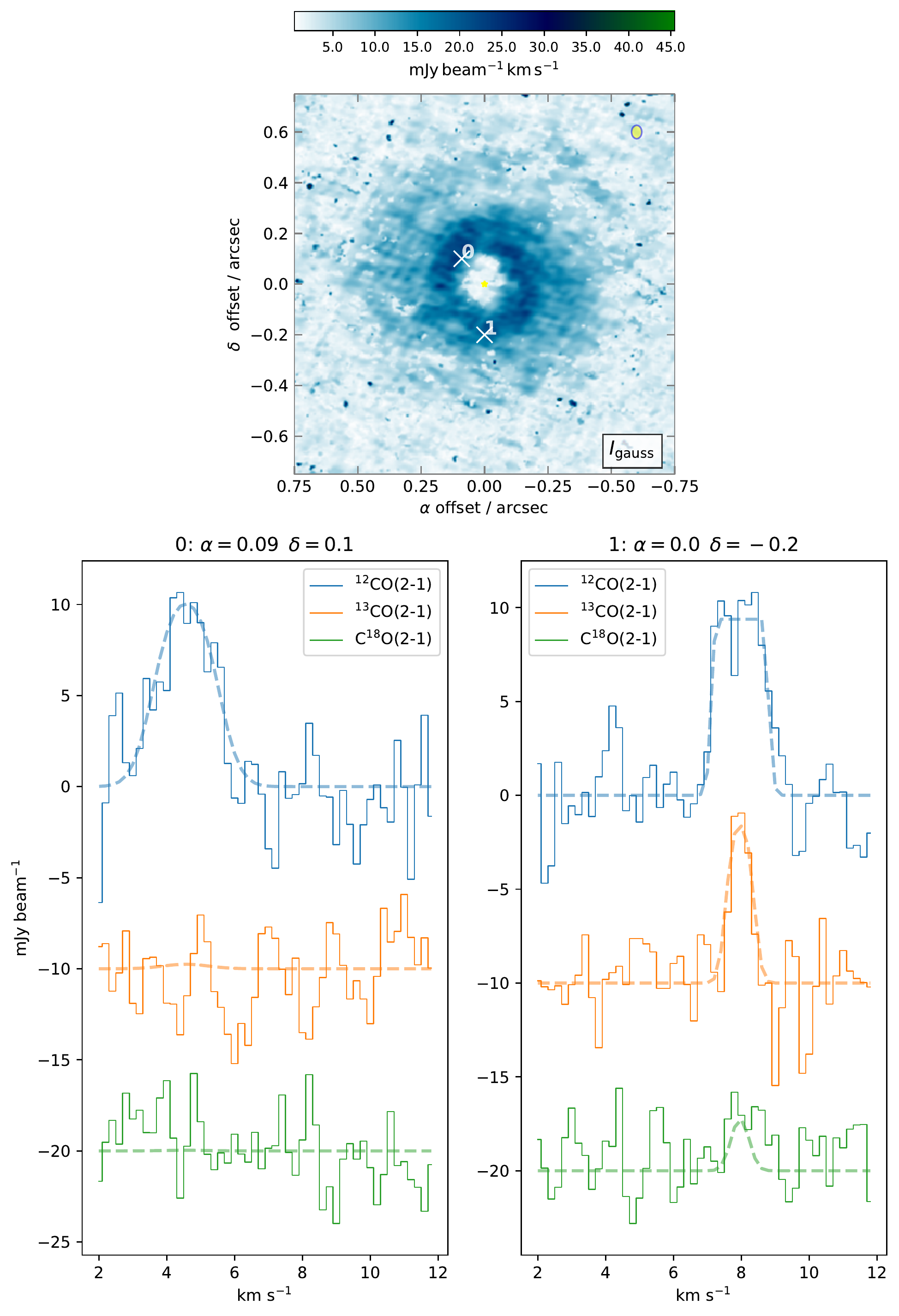}
  \caption{Comparison of observed and best fit {\sc Slab.Line} spectra
    for selected lines of sights, at the offset R.A. ($\alpha$) and
    Dec. ($\delta$) reported as titles to the spectral profiles, and
    also shown in the image at the top. The observed spectra are drawn
    in solid lines, while the model is shown in thick dashed
    lines. {\bf Left:} {\sc Slab.Line} fits on the $uv$-tapered
    datacubes. {\bf Right:} {\sc Slab.Line} fits on the original 
    datacubes (no taper).  } \label{fig:Slab_examplelos}
\end{figure*}

As explained in Sec.\,\ref{sec:lines}, the uniform-slab parameters for
the physical conditions along each line of sight are
$(\log_{10}(\Sigma_g), \log_{10}(T_b), v_\circ, \log_{10}(v_{\rm
  turb}))$. The complete set of parameters resulting from an
application of {\sc Slab.Line} to HD\,135344B, and their associated
uncertainties, are shown in Figs.\,\ref{fig:slabline} and
\ref{fig:slabline_uvtaper}, where we converted the logarithms to
linear quantities. For
conciseness the upwards and downwards uncertainties (corresponding to
the 16\% and 84\% quantiles) were averaged in a single 1$\sigma$ error
bar. Example fits are shown in Figs.\,\ref{fig:Slab_examplelos}.

Pure least-squares fit using Eq.\,\ref{eq:slablinechi2} sometimes yielded glitches in the
best-fit parameters in particularly noisy lines of sights.
After checking, in  all regions with clear
signal, that $v_{\rm turb}$ was subsonic and that the optical depth in thinnest
transition ($^{18}$CO(2-1)) was everywhere less than 5, we  controlled these glitches by adding two
regularisation terms. The final  
log-likelihood  is
\begin{multline}
  L = -\frac{1}{2}\chi^2 + \lambda_{v_{\rm turb}} H\left[\frac{(v_{\rm turb} - c_s)}{c_s}\right] \left( \frac{(v_{\rm turb} - c_s)}{c_s} \right)^2 + \\
  \lambda_{\tau} H\left[{\rm min}\left(\left\{\tau_{\circ,i}\right\}_{i=1}^{N_{\rm iso}}\right)  - \Gamma_\circ \right] \left( {\rm min}\left(\left\{\tau_{\circ,i}\right\}_{i=1}^{N_{\rm iso}}\right)  - \Gamma_\circ  \right)^2, \label{eq:loglikeslabline}
\end{multline}
where $\chi^2$ is given by Eq.\,\ref{eq:slablinechi2}, $H$ represents
the Heaviside step function and ${\rm
  min}({\tau_\circ,i}_{i=1}^{N_{\rm iso}})$ represents the minimum
optical depth at the line centre for the $N_{\rm iso}$ isotopologues
involved in the fit. We used $\lambda_{v_{\rm turb}} = 100$,
$\lambda_{\tau}=1$ and a threshold optical depth $\Gamma_\circ =5$.

\begin{figure*}
  \centering
  \includegraphics[width=0.7\textwidth,height=!]{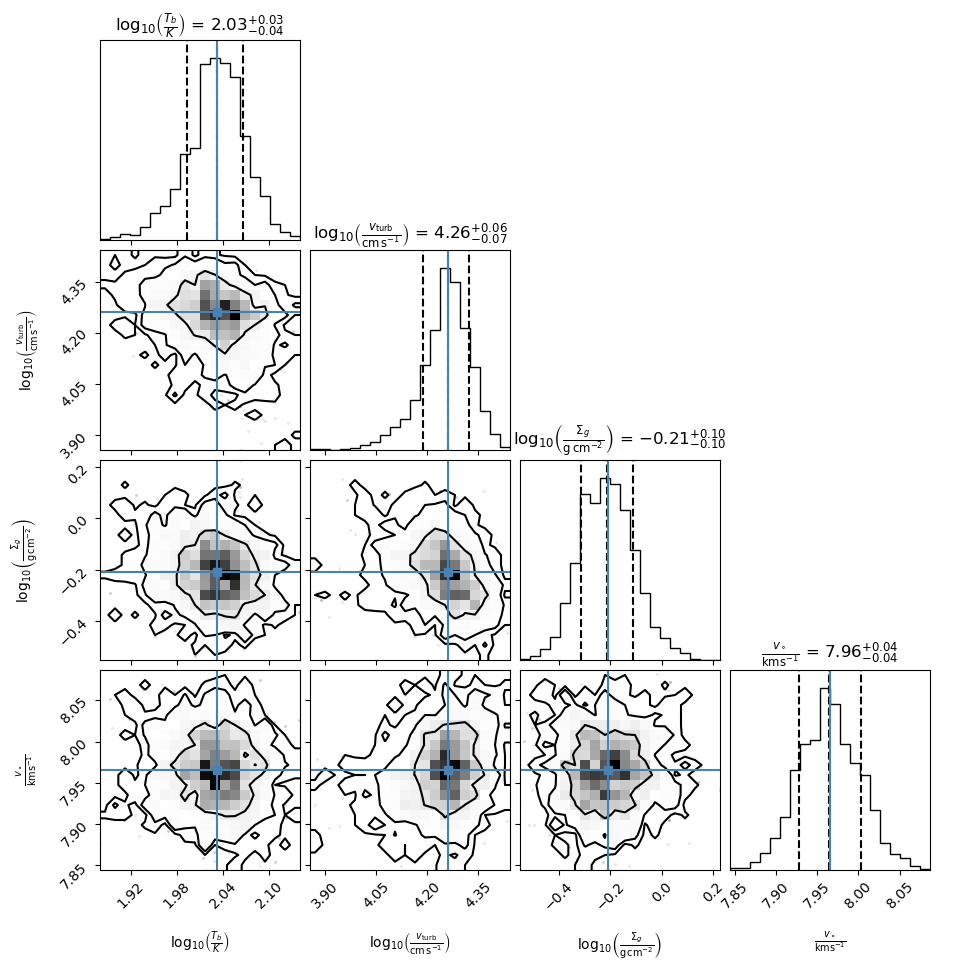}
  \caption{Corner plot for the posterior distribution of the {\sc Slab.Line} parameters in the line of sight labelled '1' in Fig.\,\ref{fig:Slab_examplelos} (right).  } \label{fig:slablinecorner}
\end{figure*}

\begin{figure*}
  \centering
  \includegraphics[width=0.7\textwidth,height=!]{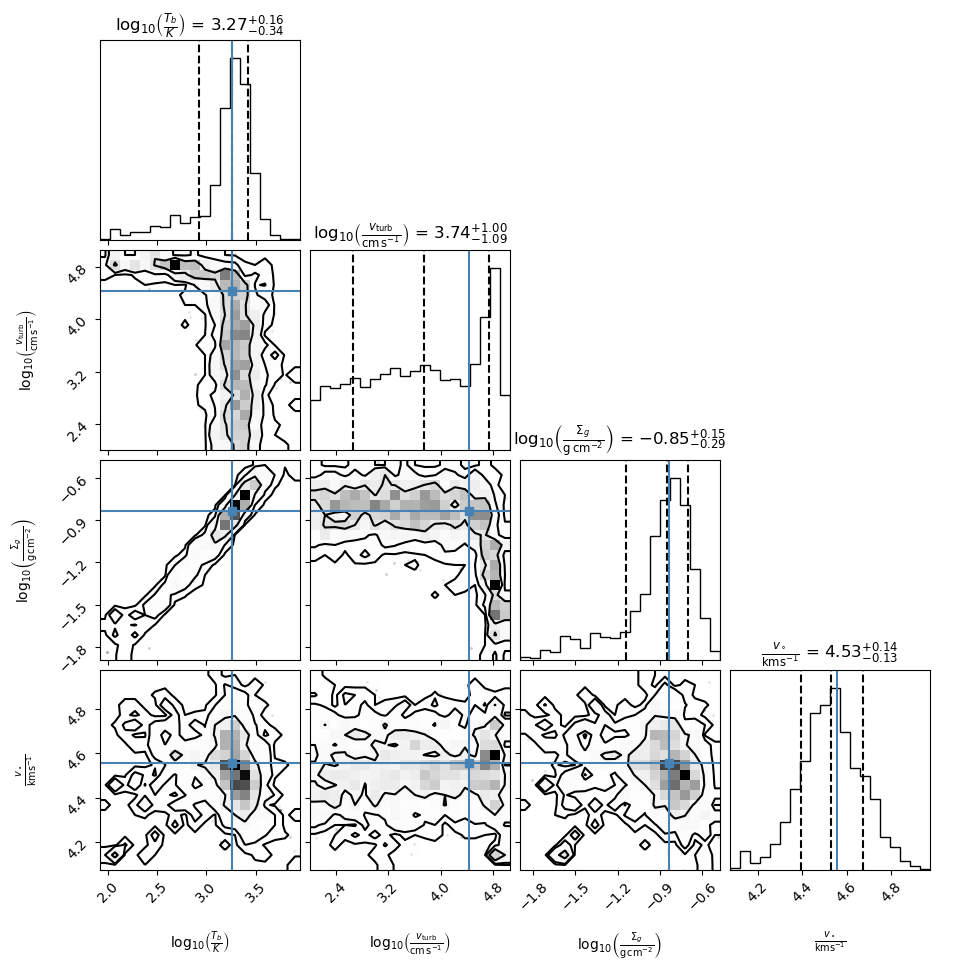}
  \caption{Same as Fig.\,\ref{fig:slablinecorner}  for the line of sight labelled '0' in Fig.\,\ref{fig:Slab_examplelos} (right).  } \label{fig:slablinecornercavity}
\end{figure*}

An example corner plot, for the line of sight labelled `1' in
Fig.\,\ref{fig:Slab_examplelos} (right, without a $uv$-taper), is
shown in Fig.\,\ref{fig:slablinecorner}. This example line of sight
has copious signal in $^{13}$CO(2-1) and leads to well-constrained
expectation values. However, the line of sight labelled `0' in
Fig.\,\ref{fig:Slab_examplelos} (right, without a $uv$-taper) falls
inside the dust ring ring and only $^{12}$CO(2-1) is picked up. The
posterior distribution of the {\sc Slab.Line} parameters are strongly
correlated, as shown in Fig.\,\ref{fig:slablinecornercavity}. The
regularisation terms used in Eq.\,\ref{eq:loglikeslabline} have no
impact in this case, and we controlled such noisy lines of sight by
imposing an upper limit temperature of 3$\times$ the $^{12}$CO(2-1)
peak brightness temperature. If instead we set an absolute maximum
temperature, for instance 500\,K, then the maps shown in
Fig.\,\ref{fig:slabline} and \ref{fig:slabline_uvtaper} are unchanged
except for $T$, which reaches somewhat higher values inside the cavity
but is modulated by noisy spikes in the outer regions where only
$^{12}$CO(2-1) is picked up.

\bsp	
\label{lastpage}
\end{document}